\documentclass[12pt]{article}
\usepackage{graphicx}
\newlength{\dummysp}
\newcommand{\beq}{\begin{equation}}
\newcommand{\eeq}{\end{equation}}
\newcommand{\ben}{\begin{enumerate}}
\newcommand{\een}{\end{enumerate}}

\newcommand{\Imag}{\mathop{{\hbox{Im} \, }}\nolimits}
\newcommand{\Real}{\mathop{{\hbox{Re} \, }}\nolimits}
\newcommand{\tr}{\mathop{{\hbox{tr} \, }}\nolimits}
\newcommand{\mtxt}[1]{\mathop{\hbox{{\small #1}}}\nolimits}

\newcommand{\third}{{1 \over 3}}

\newcommand{\beqa}{\begin{eqnarray}}
\newcommand{\eeqa}{\end{eqnarray}}
\newcommand{\nnn}{ \nonumber \\ }
\newcommand{\mod}{{\; \mtxt{mod} \; }}
\newcommand{\p}{{\partial}}
\newcommand{\Zbf}{{{\bf Z}}}

\newcommand{\Cbf}{{{\bf C}}}

\newcommand{\e}{{\epsilon}}
\newcommand{\s}{{\sigma}}

\newcommand{\vev}[1]{{\langle #1 \rangle}}

\newcommand{\ord}[1]{{{\cal O}(10^{#1})}}
\newcommand{\ordnt}[1]{{{\cal O}(#1)}}

\newcommand{\gappeq}{\mathrel{\rlap {\raise.5ex\hbox{$>$}}
{\lower.5ex\hbox{$\sim$}}}}
\newcommand{\lappeq}{\mathrel{\rlap{\raise.5ex\hbox{$<$}}
{\lower.5ex\hbox{$\sim$}}}}
\newcommand{\myref}[1]{(\ref{#1})}
\newcommand{\ux}{$U(1)_X$}
\newcommand{\dx}[1]{\p X^{#1}}
\newcommand{\dxb}[1]{\bar \p \bar X^{#1}}

\newcommand{\cab}{{\theta}}
\newcommand{\bari}{{\bar \imath}}
\DeclareFontFamily{OMS}{rsfs}{\skewchar\font'177}
\DeclareFontShape{OMS}{rsfs}{m}{n}{%
      <5> rsfs5
      <6> <7> rsfs7
      <8> <9> <10> rsfs10
      <10.95> <12> <14.4> <17.28> <20.74> <24.88> rsfs10
      }{}
\DeclareSymbolFont{rsfscript}{OMS}{rsfs}{m}{n}
\DeclareSymbolFontAlphabet{\mathrsfs}{rsfscript}
\newcommand{\mysection}[1]{\section{#1}
   \hspace{0.8cm}\setcounter{equation}{0}}
\renewcommand{\theequation}{\arabic{section}.\arabic{equation}}
\newcommand{\myappendix}{\appendix
   \renewcommand{\theequation}{\Alph{section}.\arabic{equation}}
   \vspace{30pt} \noindent {\Large \bf Appendix}}
\begin{document}

\begin{titlepage} 

\baselineskip=14pt

\hfill    LBNL-46292

\hfill    UCB-PTH-00/22

\hfill    hep-ph/0007193

\hfill    Dec.~17, 2001

\begin{center}

\vspace{20pt}

  { \bf \Large The KM phase in semi-realistic \\
\vspace{5pt}
   heterotic orbifold models}

\end{center}

\begin{center}
{\sl Joel Giedt${}^*$}

\end{center}

\begin{center}

{\it Department of Physics, University of California, \\
and Theoretical Physics Group, 50A-5101, \\
Lawrence Berkeley National Laboratory, Berkeley,
CA 94720 USA.}\footnote{This work was supported in part by the
Director, Office of Science, Office of High Energy and Nuclear
Physics, Division of High Energy Physics of the U.S. Department of
Energy under Contract DE-AC03-76SF00098 and in part by the National
Science Foundation under grant PHY-95-14797.}

\end{center}

\begin{center}

{\bf Abstract}

\end{center}

\vspace{5pt}

In string-inspired semi-realistic heterotic
orbifolds models with an anomalous $U(1)_X$,
a nonzero Kobayashi-Masakawa (KM) phase is 
shown to arise generically 
from the expectation values of 
complex scalar fields, which appear
in nonrenormalizable quark mass couplings.
Modular covariant nonrenormalizable
superpotential couplings are constructed.
A toy $Z_3$ orbifold model is
analyzed in some detail.  Modular 
symmetries and orbifold selection rules 
are taken into account and do
not lead to a cancellation 
of the KM phase. 
We also discuss attempts to obtain
the KM phase solely from renormalizable
interactions.

\vfill

\begin{tabbing}

{}~~~~~~~~~\= blah  \kill
\> ${}^*$ E-Mail: {\tt JTGiedt@lbl.gov}

\end{tabbing}

\end{titlepage}

\renewcommand{\thepage}{\roman{page}}
\setcounter{page}{2}
\mbox{ }

\vskip 1in

\begin{center}
{\bf Disclaimer}
\end{center}

\vskip .2in

\begin{scriptsize}
\begin{quotation}
This document was prepared as an account of work sponsored by the United
States Government. Neither the United States Government nor any agency
thereof, nor The Regents of the University of California, nor any of their
employees, makes any warranty, express or implied, or assumes any legal
liability or responsibility for the accuracy, completeness, or usefulness
of any information, apparatus, product, or process disclosed, or represents
that its use would not infringe privately owned rights. Reference herein
to any specific commercial products process, or service by its trade name,
trademark, manufacturer, or otherwise, does not necessarily constitute or
imply its endorsement, recommendation, or favoring by the United States
Government or any agency thereof, or The Regents of the University of
California. The views and opinions of authors expressed herein do not
necessarily state or reflect those of the United States Government or any
agency thereof of The Regents of the University of California and shall
not be used for advertising or product endorsement purposes.
\end{quotation}
\end{scriptsize}

\vskip 2in

\begin{center}
\begin{small}
{\it Lawrence Berkeley Laboratory is an equal opportunity employer.}
\end{small}
\end{center}

\newpage
\renewcommand{\thepage}{\arabic{page}}
\setcounter{page}{1}
\def\thefootnote{\arabic{footnote}}
\setcounter{footnote}{0}

\baselineskip=16pt

%
%

\mysection{Introduction}
\label{intro}
Numerous attempts have been made
to explain the origin of CP violation
in the context of string-derived effective
supergravity models.  Early on, Strominger and Witten
suggested compactification might lead to {\it explicit}
violation of CP from the four-dimensional
point of view, since the operation of 
CP is orientation-changing
in the six-dimensional compact space \cite{strom:85a}.
Inspired by this prospect, 
detailed analyses were carried out
for various compactifications of the heterotic string;
for example, a Calabi-Yau manifold was
examined in \cite{limcs:91a} and $Z_N$ 
orbifolds were investigated in \cite{kobay:95a}.
In each case where explicit
CP violation was investigated, it was
found not to occur.  Subsequently,
Dine et al.~argued that CP is 
a gauge symmetry in string theory
and that explicit breaking is therefore forbidden both
perturbatively and nonperturbatively \cite{dinel:92a}.
In the same work it was shown that this is certainly
true for heterotic orbifolds.
Based on these results,
it is clear that if a heterotic 
orbifold is to provide a reasonable approximation
to the correct underlying theory of known interactions,
CP violation must occur {\it spontaneously} from
string moduli or matter fields (or perhaps both)
acquiring vacuum expectation values ({\it vevs}).

In a series of papers, Bailin,
Kraniotis and Love (BKL) considered
the possibility of supersymmetric CP
violation (SCPV)\footnote{By this, we mean
CP-violating complex phases in the
soft supersymmetry breaking operators of minimal extensions to 
the Standard Model (SM) \cite{scpv}.}
in heterotic orbifold models \cite{bkl}. 
They related SCPV to the complex phases of string moduli vevs.
While SCPV is an interesting possibility, 
it is important to understand
how the Kobayashi-Masakawa (KM) 
phase \cite{kobay:73a}
might occur in semi-realistic
orbifold models and how generic it is.
Furthermore, SCPV in the context of
supergravity is phenomenologically problematic
unless soft terms meet a variety of stringent
criteria \cite{scpv,dugan:85a}; the current
status of SCPV and various solutions
to related phenomenological problems are
reviewed in ref.~\cite{gross:97a}.
By contrast, the KM phase does not have
dangerous side-effects; the smallness
of CP violation is for the most part explained
by the small mixing angles between heavy and
light generation quarks while large
electric dipole moments do not arise
from this source \cite{gavel:82a}.

BKL have considered complex vevs for
the string moduli of heterotic orbifolds
as a source of the KM phase \cite{bklckm}.
{\it T-} and {\it U-moduli,} which parameterize
deformations of the six-dimensional compact
space consistent with the orbifold construction,
enter into the effective Yukawa
couplings for twisted fields, as
described below.  In this article
we will restrict
ourselves to $Z_3$ and $Z_3 \times Z_3$
orbifolds, which have
no U-moduli because the complex structure
is fixed.  It has been demonstrated
by BKL and others that is possible for nonperturbative
effects to stabilize these string moduli at
complex values \cite{bkl,binet:97a}.  BKL have
shown that, as a result of complex string moduli vevs, 
$\ordnt{1}$ phases can arise in twisted Yukawa
coupling coefficients.  However, they do not construct
any models and it is unclear whether or 
not the phases they find are physical:  
phases in the quark Yukawa matrices
do not necessarily imply the existence
of a nonzero KM phase.\footnote{Complex  
phases which give rise to a nontrivial KM
phase will be termed {\it physical} 
while those which can be eliminated by
rephasing quark fields will be termed {\it spurious}.}
As an example,
let us write the quark mass superpotential as
\beq
W_{qm} = \lambda_{ij}^u H_u Q_{iL} u_{jL}^c
+ \lambda_{ij}^d H_d Q_{iL} d_{jL}^c
\label{sqm}
\eeq
and suppose Yukawa matrices of the form
\beq
\lambda^u_{ij} = |\lambda^u_{ij}| e^{i(\alpha_i+\beta^u_j)},
\qquad 
\lambda^d_{ij} = |\lambda^d_{ij}| e^{i(\alpha_i+\beta^d_j)}.
\label{exy}
\eeq
The phases $\alpha_i,\beta^{u,d}_j$ are not physical
since they can be removed by rephasing the quark fields
according to
\beq
Q_{iL} \to e^{-i\alpha_i} Q_{iL},
\qquad u_{jL}^c \to e^{-i\beta^u_j} u_{jL}^c,
\qquad d_{jL}^c \to e^{-i\beta^d_j} d_{jL}^c.
\eeq
Naive orbifold models of quark Yukawa couplings
assign the nine SM multiplets 
$Q_{iL},u_{jL}^c,d_{jL}^c$ to different
sectors of the Hilbert space and
rely on trilinear couplings to give
all of the quarks mass; the assignments
to different sectors 
determine the moduli-dependence of the
effective Yukawa matrices, hence
the complex phases.  If these assignments can be brought
into correspondence with phases that enter the
Yukawa matrices in a way similar to \myref{exy},
then they will not give rise to CP violation.
An example of this is described in Section \ref{bklckm_sec} below.
Furthermore, when one reviews the tables of
phases displayed in the appendix
of ref.~\cite{bklckm}, one finds that many of
them are identical or zero for a given
orbifold model and T-modulus vev.  It then
becomes a concern whether or not this high
degree of degeneracy causes the
phases to ``wash out'' in the final
analysis.  An example of this is also given
in Section~\ref{bklckm_sec}.

Beyond these more obvious ways that complex
phases in the Yukawa matrices may not give
rise to a nontrivial KM phase, we must also be concerned that
symmetry constraints imposed on the Yukawa
matrices by the underlying string theory 
might relate the phases to each other
in such a way that the CKM matrix can be
made entirely real.  The embedding of the
orbifold action into the internal left-moving
gauge degrees of freedom
typically leaves a surviving
gauge group significantly larger than the
SM.  For example, in the three generation constructions
which we will discuss below the surviving
gauge group $G$ has rank sixteen \cite{thrgen,font:90a}.
Low-energy effective Yukawa couplings
must be constructed from high-energy operators
invariant under $G$.
The high-energy operators are also subject
to orbifold selection rules,
which result from symmetries of
the six-dimensional compact space.  
(A brief review of
orbifold selection rules may be found in ref.~\cite{font:90a}.)
Finally, the underlying
conformal field theory has target space
modular symmetries associated with the
identification of equivalent string moduli backgrounds.
It is conceivable (albeit highly unlikely) that orbifold selection
rules, gauge invariance under $G$ 
and target-space modular invariance may conspire to
make phases derived from the 
scalar background spurious.
In the present
paper we construct a toy $Z_3$ orbifold
model which is subject to these symmetry
constraints; we construct
explicit quark mass matrices in order
to make conclusive statements about the
existence of a nonzero KM phase.

The case of complex vevs for the
T-moduli is treated below; however, we place
more emphasis on another origin of complex
phases, which in our opinion is a much
more generic and natural source of CP violation
in semi-realistic heterotic orbifold models.
Nonrenormalizable couplings are often
important in semi-realistic heterotic
orbifold models for the following reason.
Wilson lines are typically included in the
embedding to get a reasonable gauge group.\footnote{
Standard GUT scenarios require large higgs
representations, whereas they are absent in 
affine level 1 orbifold constructions \cite{gutorb}.
This makes it phenomenologically advantageous to obtain
$G=G_{SM} \times G_{other}$ from the start.}
In many cases their inclusion leads to
an anomalous $U(1)_X$ factor:  $\tr Q_X \not= 0$.
The apparent anomaly is cancelled 
by the Green-Schwarz
mechanism, which induces a Fayet-Illiopoulos (FI)
term \cite{dine:87}.  
Several scalar fields get vevs $v_i \sim \ord{-2\pm 1}$ 
(in units where $m_P = 1 / \sqrt{8 \pi G} = 1$)
to restabilize the vacuum in the presence of the FI term;
thus, the suppression of nonrenormalizable couplings
due to vevs $v_i$ may be as little as $\ord{-1}$.
The scalar fields which cancel the FI term
break the \ux\ gauge symmetry at
the scale $\Lambda_X \sim 10^{17}$ GeV by
the Higgs mechanism; in order to distinguish
them from the higgses associated with 
electroweak symmetry breaking, we will
for convenience and with all due apologies
refer to them as {\it Xiggses.}
The Xiggses are usually charged under other
$U(1)$ factors of $G$ besides $U(1)_X$;
a basis of generators can always be chosen
such that these other $U(1)$'s are not
anomalous.\footnote{We will not consider  
the case where Xiggses are in nontrivial
representations of the nonabelian factors
of $G$.}  Vacuum stabilization near
the scale $\Lambda_X$ requires 
the D-terms of these other $U(1)$'s to vanish.
To satisfy the numerous D-flatness conditions,
it is generally necessary for several Xiggses to get
vevs.  Typically there are $\ordnt{10}$ 
or more such Xiggses.  

In this paper we demonstrate how
the KM phase in semi-realistic heterotic orbifold models
arises generically from the ``Planck slop''
created by the Xiggses. 
Nonrenormalizable couplings make significant and in
some cases leading order contributions 
to the effective quark mass matrices.  For instance,
in the semi-realistic model developed
by Font, Ib\'a\~nez, Quevedo and Sierra in 
Section 4.2 of ref.~\cite{font:90a}
({\it FIQS model}), 
the down-type quarks get their leading order mass from 
dimension 10 holomorphic couplings.  (We count dimensions
as 1 for each elementary superfield entering a coupling.)
For up-type quarks, only
the top and charm get masses from renormalizable 
couplings.\footnote{We identify top, charm, etc., by
ranking the mass at a given level of analysis;
of course, the identification will be imperfect
once higher-order corrections are included.}
The up quark must receive its mass from nonrenormalizable couplings
or radiative mass terms. Although high order 
nonrenormalizable couplings
are suppressed by large powers of the $\ord{-2\pm 1}$
Xiggs vevs, the suppression is not as large
as one might think, for two reasons.
Many high order operators
involving the Xiggses generically exist,
with their number increasing
at each higher order.
Most of the operators are closely related by variations in fixed
point locations and oscillator ``directions'', as will be shown
in detail below; however, this does not change the fact
that the number of distinct operators tends to be large.
The second reason why high order couplings
are important is that the coupling strength tends to
be much larger than $\ordnt{1}$ and to grow
as the dimension of the coupling increases, as was pointed
out by Cveti\v c et al.~\cite{cvetic:98}.
The combination of these two effects forces one to
proceed to rather high order before couplings
make negligible contributions to the quark mass
matrices.  Here, ``negligible'' is taken to mean
less than, say, 10\% contributions to the lightest
quark masses. 
As a consequence, each effective Yukawa coupling
$\lambda_{ij}^{u,d}$ depends on the vev 
of a linear combination of a large number of 
monomials of Xiggses.  The principal point of
this article is that since $\ordnt{10}$ Xiggses get
complex vevs, which appear in a large
number of monomials contributing to effective
quark Yukawa couplings, the 
Yukawa matrices are generically complex
and a nonvanishing KM phase is almost inevitable.
Indeed, in any orbifold model with an
anomalous \ux\ present, it seems improbable
that one would not have CP violation
in this way, since it is difficult to see
how nonrenormalizable couplings involving the
Xiggses would not contribute to the effective
quark Yukawa couplings.  

Previous authors have noted the possible role
of nonrenormalizable couplings in heterotic orbifold
and free fermionic models for giving large mass hierachies and
CP violation from a KM phase \cite{kx,blrvw,farag:94a}.
In this respect the mechanism analyzed
here is not new.  However, we present much
more detailed results by constructing an
explicit toy model and we impose modular
invariance on the nonrenormalizable couplings.
The constraint of modular invariance
leads to relationships between coupling
coefficients which were not accounted for
in earlier efforts.  The model is inspired
by three generation heterotic $Z_3$ orbifold models
previously investigated in the literature
\cite{thrgen, font:90a}.  Our model
is, by construction, quite similar to
the FIQS model, which is string-derived:
in the FIQS model, the gauge group, 
the spectrum of states
and the allowed superpotential couplings 
are completely determined from
the underlying string theory.
The FIQS model suffers from
phenomenolgical difficulties
related to the quark mass matrices; these difficulties
were previously pointed out in \cite{gaill:00a}.
At leading order in the FIQS model, the top and charm
come from different $SU(2)$ doublets
than the bottom.  As a result, the
leading order CKM matrix has some of its
diagonal entries zero and some $\ordnt{1}$ heavy-light
generation mixing angles, which is clearly
unacceptable.  The assignments into $SU(2)$
doublets are determined by the H-momenta\footnote{
H-momenta are the $SO(10)$ weights of
bosonized NSR fermions, which appear
in the vertex operators creating
asymptotic states.}
of the untwisted states, so the identification
of which doublet a given $u_{iL}$ or $d_{iL}$
sits in and how it couples at leading
order is fixed.  A second problem with
the FIQS model is that it is difficult to
give the three lightest quarks mass.
We have searched for dimension
$d \le 30$ allowed 
holomorphic couplings which might do the
job (under the assumptions made in the FIQS
model about Xiggs and T-moduli vevs)
and found that none exist.
We further found that radiative masses
only occur at high loop levels and would be
minuscule in comparison to the experimental
values.  However, we assumed
the leading order K\"ahler potential in
this analysis.  It is possible that
higher order terms in the K\"ahler potential
could provide a mechanism for giving the
light quarks masses in agreement
with experimental values.

We have attempted in many ways to evade these
problems.  However, we were not able to do so without
creating other difficulties.  We are actively
searching for a superior three generation heterotic
$Z_3$ orbifold model to study.  In the meantime, we have
developed a toy model which replicates the FIQS model
wherever possible while avoiding its problems,
in order that we might illustrate
that a nontrivial KM phase generically
arises from the complex vevs of Xiggses,
even after orbifold selection rules
and target space modular invariance have
been accounted for.

Although the coupling coefficients for
nonrenormalizable superpotential
couplings (which are in principle
obtainable from the underlying conformal field
theory) are not apparently known, we propose
couplings which transform in the requisite manner
under the $[SL(2,\Zbf)]^3$ diagonal subgroup
of the full $SU(3,3,\Zbf)$ modular duality group
of the $Z_3$ orbifold \cite{z3dual}.
We take into account the non-trivial
transformations of twisted sector fields in
our construction of modular invariant couplings.
In the three generation $Z_3$ orbifold models upon
which the toy model of Section \ref{pckm} is based,
different {\it species}
are distiguished by quantum numbers (other
than the fixed point location in the third
complex plane) of
massless states in the underlying theory \cite{font:90a,thrgen}.
For twisted fields $\Phi_n^i$, where $n$ labels the
species and $i=1,2,3$ labels the
fixed point locations in the third complex
plane, $\Phi_n^1, \Phi_n^2$ and $\Phi_n^3$ 
mix amongst themselves
under the $T^3 \to 1/T^3$ duality transformation
in the third complex plane \cite{chun:89,lauer:91a}.
Constraining holomorphic polynomials of dimension $d>3$
to transform with\footnote{Modular weight
will be explained below.}  
modular weight $-1$ in light
of these nontrivial mixings places strong
constraints on the form of the superpotential
terms and gives some confidence that our
proposed nonrenormalizable
coupling coefficients may reproduce key features of the
actual couplings which would be
derived using conformal field theory techniques.
In the course of discussing our assumptions
for the coupling coefficients of nonrenormalizable
superpotential terms, we will explain
why the calculation of these coefficients from the
underlying string theory represents an
extremely difficult problem.  For
now, we remark that the
most intimidating aspect of such a calculation
is the integration of the string correlator
over the $d-3$ vertex locations which cannot
be fixed by $SL(2,\Cbf)$ invariance,
where $d$ is the dimension of the coupling.
The string correlator is generally a very
complicated function of the unfixed vertex
locations.

The reliability of the effective supergravity
approach is hampered
by theoretical uncertainties in the K\"ahler
potential of $Z_3$ orbifold models.  
Nonleading operators in
the K\"ahler potential are neglected in most analyses;
however, with $\ord{-2\pm 1}$ Xiggs vevs,
higher order terms in the
K\"ahler potential give non-negligible corrections
to the mass matrices of quarks:  corrections
from higher order terms give the quarks noncanonical,
nondiagonal kinetic terms which must
be rendered canonical by nonunitary field redefinitions
when one goes to compute mass eigenstates and mixings.
These complications cannot
be ignored if one hopes to develop an accurate
picture of the low-energy phenomenology predicted
by a given model.
Higher order terms in the K\"ahler
potential ought to be included in order to
be consistent with the high order expansion of
the superpotential, both of which are
necessary in order to pick up all
significant contributions to the quark
mass matrices.
The calculation
of higher order corrections to the K\"ahler potential
is notoriously difficult because of
the lack of holomorphicity.  
It is hoped that
future work on the K\"ahler potential of
heterotic $Z_3$ orbifold models will
amend these deficiencies and
allow for an improved analysis of the low energy
phenomenology of semi-realistic models.
Present ignorance
regarding these aspects of the K\"ahler potential
has forced us to make a number of oversimplifications.
However, we do not expect these oversimplifications
to affect our main result that a nontrivial KM phase is generic.
Introducing higher order terms in the K\"ahler potential
is not expected to eliminate the complex phases
which enter into the effective quark Yukawa couplings from
the vevs of Xiggses.

In Section \ref{bklckm_sec} we present examples
where the complex phases found by BKL are spurious.
In Section \ref{modinv}
we introduce modular invariant coupling coefficients
for nonrenormalizable superpotential couplings.
In Section \ref{pckm} we discuss our string-inspired
toy heterotic $Z_3$ orbifold model.
In Section \ref{concl} we make concluding remarks
and suggest further investigations 
motivated by our results.
In the Appendix we address normalization
conventions for $U(1)$ charges
in string-derived models.
We explain how to account for different
conventions when determining the FI term and
describe the Green-Schwarz cancellation of the \ux\
anomaly in the linear multiplet formulation.

%
%

\mysection{Counterexample}
\label{bklckm_sec}
Here, we consider some assignments of quarks and higgses
in $Z_3 \times Z_3$ orbifold models and show that the
complex phases found by BKL do not
lead to a nontrivial KM phase for these particular
examples.  We certainly do not wish to imply that the
phases found by these authors cannot lead
to a nonzero KM phase; we only wish to point out
that due to the degeneracy in phases
(in this case always $0$ or $-\pi/3$),
they can in many cases
be eliminated by rephasing quark fields.
In our opinion, a more
careful analysis is required in order to conclude
whether or not the phases
found by BKL can account
for CP violation.

We will use the notation and conventions of
ref.~\cite{bls} in our discussion of the
twisted sectors and fixed tori of the $Z_3
\times Z_3$ orbifold.  This orbifold is
constructed using twists
\beq
\theta = \third (1,0,-1), 
\qquad \omega = \third (0,1,-1).
\eeq
We make use of the
$\theta$, $\theta \omega$ and $\theta \omega^2$
twisted sectors.  The fixed tori for each of
these sectors are given by
\beqa
f_\theta & = & {m_1 \over 3}(2 e_1 + \tilde e_1)
+ {m_3 \over 3}(e_3 - \tilde e_3) + v_2, 
\; m_{1,3}=0, \pm 1, 
\; v_2 \in K_2, \label{fpt} \\
f_{\theta \omega} & = & {1 \over 3}
\sum_{i=1}^3 r_i (2 e_i + \tilde e_i) + \ell,
\quad r_i = 0, \pm 1,
\quad \ell \in \Lambda, \\
f_{\theta \omega^2} & = & {p_1 \over 3}(2 e_1 + \tilde e_1)
+ {p_2 \over 3}(e_2 - \tilde e_2) + v_3,
\; p_{1,2} = 0, \pm 1,
\; v_3 \in K_3, \label{fptw2}
\eeqa
where $\Lambda$ is the
$[SU(3)]^3$ root lattice and $K_i$ is
the $i$th complex plane.
Physical states
must be simultaneous eigenstates of $\theta$
and $\omega$; they are therefore linear
combinations of states whose zero modes are
given by different fixed tori.
Since the first complex
plane is neutral under $\omega$, physical
states in the $\theta$ sector can be chosen
with a definite quantum number $m_1$.  Because the
form of the fixed torus \myref{fptw2} in
the first complex plane is the same as in
the $\theta$ sector \myref{fpt}, and since
$\omega$ does not rotate in the first plane,
a physical state in the $\theta \omega^2$
sector can be chosen to have a definite
quantum number $p_1$ as well.  
The coefficients of the trilinear
Yukawa couplings are determined
by the evaluation of correlation
functions in the underlying string theory.
The contribution from a complex $T^1$ 
in the classical
partition function is the only source
of complex phases in the trilinear
couplings.  The phases
of $T^{2,3}$ do not matter because
the twist operator contributions to the
correlation function in the second and
third complex planes reduce to the identity,
as discussed in ref.~\cite{bls}.
The phases found in ref.~\cite{bklckm} for the case
$\vev{T^1}= \exp (i \pi / 6)$ are determined by the
difference $m_1-p_1$:
\beq
\gamma(m_1,p_1) \equiv
\arg \sum_{X_{cl}} e^{-S_{cl}} 
=
\left\{
\begin{array}{ll}
0, & m_1-p_1 = 0; \\
- {\pi \over 3}, & m_1-p_1 = \pm 1, \pm 2.
\end{array}
\right.
\label{bph}
\eeq

We next suppose assignments as follows:
$H_u,H_d$ are in the $\theta$ sector
with fixed tori quantum numbers $m_1^u,m_1^d$ resp.;
$Q_{iL}$ are in the $\theta \omega^2$
sector with fixed tori quantum numbers
$p_1^i$ resp.; $u_{jL}^c, d_{jL}^c$
are in the $\theta \omega$ sector.
The phases \myref{bph} enter into the
Yukawa couplings \myref{sqm} according to:
\beq
\lambda^u_{ij} = |\lambda^u_{ij}| e^{i\alpha^u_i}, \qquad
\lambda^d_{ij} = |\lambda^d_{ij}| e^{i\alpha^d_i},
\eeq
where $\alpha^u_i=\gamma(m_1^u,p_1^i)$ and
$\alpha^d_i=\gamma(m_1^d,p_1^i)$.  
If $m_1^u=m_1^d$
then we recover the CP conserving forms
of \myref{exy} with $\alpha_i = \alpha^u_i=\alpha^d_i$
and $\beta^{u,d}_j=0$.  

It is an interesting question
whether or not CP violation really can occur,
and the example just presented demonstrates
that a careful case-by-case analysis
is probably required in order to draw
any firm conclusions.  It should be noted
that in the case where nonrenormalizable
couplings give significant corrections to
the trilinear $(\theta,\theta \omega,\theta \omega^2)$
coupling, the above arguments likely fail.

Lastly, we would like to comment on another suggested
source of complex phases.  As noted above,
physical states are constructed
from linear combinations of states whose zero modes
are given by different fixed tori.
It was noted by Kobayashi and Ohtsubo
that complex phases enter from the
coefficients in the linear combinations
and that these might be a souce of CP
violation \cite{ko:90}.  However, this
amounts to explicit CP violation
since it does not require a particular
scalar background.  As explained in
Section \ref{intro}, explicit CP violation
does not occur in heterotic orbifolds
since CP is a gauge symmetry of the underlying theory.
Therefore, the phases which enter into
couplings from this source
cannot contribute to the KM phase.

%
%
%
%

\mysection{Modular invariance}
\label{modinv}
In the toy model to be considered below,
nonrenormalizable superpotential couplings
will play a crucial role.  In this section
we present a set of assumptions for
coupling coefficients of
holomorphic couplings of arbitrary
order; the result will be couplings
which transform in the requisite manner
under the $[SL(2,\Zbf)]^3$ subgroup of
the full $SU(3,3,\Zbf)$ modular duality
group of the $Z_3$ orbifold.

We begin by considering
the simpler case of a two dimensional $Z_3$ orbifold,
where there is a single modulus $T$ and there are only
three fixed points.  Consequently,
twisted matter fields carry a single fixed point
label.  The twisted trilinear couplings
are known in this simple case \cite{dixon:87}.
These couplings can
be expressed in terms of the Dedekind $\eta$ function
\beq
\eta(T) = e^{-\pi T / 12} \prod_{n=1}^\infty (1 - e^{-2 \pi n T})
\eeq
and the level-one $SU(3)$ characters \cite{chun:89}
\beq
\chi_i(T) = \eta^{-2}(T) \sum_{v \in \Gamma_i} e^{- \pi T |v|^2}.
\label{charsu3}
\eeq
In this expression, $\Gamma_0$ is the $SU(3)$ root lattice,
while $\Gamma_{1,2}$ are shifted by $SU(3)$ weight
vectors.  Explicitly,
\beq
|v|^2 = \pmatrix{n_1 + {i\over 3}, & n_2+{i\over 3} \cr}
\pmatrix{2 & -1 \cr -1 & 2 \cr}
\pmatrix{n_1+ i / 3 \cr n_2+ i/ 3 \cr}
\eeq
with $n_1$ and $n_2$ integers to be summed over
in \myref{charsu3} and $i$ is the integer labeling $\Gamma_i$.
It is easy to check that $\chi_1=\chi_2$.
The values of these three functions 
at the self-dual points $T=1,e^{i \pi / 6}$ are approximately
given by Table \ref{ecv}.
\begin{table}
\begin{center}
$$
\begin{array}{|l||l|l|l|} \hline
T & \eta(T) & \chi_0(T) & \chi_1(T) \\ \hline \hline
1 & 0.76823 & 1.7134 & 0.97280 \\ \hline
e^{i \pi / 6} & 0.80058 \, e^{-i \pi/24} 
& 1.5197 \, e^{i \pi / 12} & 0.91251 \, e^{- i \pi /12} \\ \hline
\end{array}
$$
\end{center}
\caption{Values of modular functions at self-dual points.}
\label{ecv}
\end{table}
The trilinear couplings between twisted fields
$\Phi_1^{i_1}, \Phi_2^{i_2}, \Phi_3^{i_3}$
in the superpotential are
\beq
\lambda \cdot [\eta(T)]^2 \, f_{i_1 i_2 i_3}^T \,
\, \Phi_1^{i_1} \, \Phi_2^{i_2} \, \Phi_3^{i_3},
\label{thetri}
\eeq
where $f_{i_1 i_2 i_3}^T$ is given by:
\beq
f_{i_1 i_2 i_3}^T = \left\{
\begin{array}{l}
\chi_0(T), \quad i_1=i_2=i_3; \\
\chi_1(T), \quad i_1 \not= i_2\not = i_3\not = i_1; \\
0, \quad \mtxt{else}.
\end{array} \right.
\label{dff}
\eeq
The overall T-independent
coupling strength $\lambda$ does not
depend on fixed point locations and
is obtained by factorization of the
four-point string correlator \cite{dixon:87}.
Note that $\eta(T)$ is superfluous
and occurs only because of the definition
of $\chi_i (T)$.

It can be checked that the above Yukawa couplings
transform correctly under the
target-space modular duality group $SL(2,\Zbf)$.
The K\"ahler potential for the modulus $T$ of
the 2-d orbifold is given by $K(T,\bar T)= -\ln (T + \bar T)$,
which transforms under the $SL(2,\Zbf)$ modular transformations
\beq
T \to {aT-ib \over icT+d}, \qquad ad-bc=1, \qquad a,b,c,d \in \Zbf,
\label{sl2z}
\eeq
as
\beq
K(T,\bar T) \to K(T,\bar T) + F(T) + \bar F(\bar T),
\eeq
where $F(T) = \ln (icT+d)$.
We require that the quantity $K + \ln |W|^2$ remain
invariant \cite{cremm:78a}, which implies that
the superpotential $W$ transform as
\beq
W \to e^{i\gamma(a,b,c,d)} e^{-F(T)} W, 
\label{wcf}
\eeq
where $\gamma(a,b,c,d)$ is a T-independent 
phase which does not appear
in the functional which is physically
meaningful, $K + \ln |W|^2$.
We can take $b = -1, c=1,d=0$
to obtain the $T \to 1/T$ transformation of $SL(2,\Zbf)$,
so that $F(T) = \ln (iT)$ in this case.
It has been shown \cite{chun:89,lauer:91a}
that the twist fields $\s_i$ which create twisted vacua corresponding
to fixed points labeled by $i$, and hence the twisted states $\Phi^i$,
transform under the duality transformation $T \to 1/T$ as
\beq
\pmatrix{\s_1' \cr \s_2' \cr \s_3' \cr}
= {e^{i \beta} \over \sqrt{3}}
\pmatrix{1 & 1 & 1 \cr 1 & \bar \alpha & \alpha \cr
1 & \alpha & \bar \alpha \cr}
\pmatrix{\s_1 \cr \s_2 \cr \s_3 \cr},
\eeq
where $\exp(3i\beta) \equiv \sqrt{\bar T / T}$ and $\alpha \equiv
\exp(2\pi i/3)$.  Since the vertex operator which
creates the twisted state $\Phi^i$ is proportional
to $\s_i$, the twisted fields must transform in a way
which is proportional to the same matrix.  The nonholomorphic
phase $\beta$ must be absent in the supergravity definition of
the fields.  For example, we could define 
$\tilde \s_i = e^{i \beta / 2} \s_i$
and use these to create the supergravity fields, which must transform
in a holomorphic way.  Aside from the mixing of
different fixed points, the
twisted fields tranform with a {\it modular weight} q:
$\Phi \to \Phi' \sim e^{-q F(T)} \Phi$.  These weights are
known \cite{dixon:90a,ibane:92,blmw}; a non-oscillator
twisted field has modular weight $q=2/3$.
Only non-oscillator
twisted fields are allowed to enter into the
trilinear twisted couplings due to the
{\it automorphism selection rule}, corresponding
to invariance of string correlators under
automorphisms of the underlying $SU(3)$
lattice.  This rule is explained and illustrated
with examples in ref.~\cite{font:90a}.
Since $F(T) = \ln (iT)$ for the $T \to 1/T$
duality transformation, non-oscillator twisted fields must
transform as
\beq
\pmatrix{{\Phi^1}' \cr {\Phi^2}' \cr {\Phi^3}' \cr}
= {1 \over \sqrt{3}} \left( {1 \over iT} \right)^{2/3}
\pmatrix{1 & 1 & 1 \cr 1 & \bar \alpha & \alpha \cr
1 & \alpha & \bar \alpha \cr}
\pmatrix{\Phi^1 \cr \Phi^2 \cr \Phi^3 \cr}.
\label{twtrd}
\eeq
Under the duality transformation $T \to T' =1/T$ it has been
shown \cite{chun:89} that the $SU(3)$ characters transform
as
\beq
\pmatrix{\chi_0' \cr \chi_1' \cr \chi_2' \cr}
= {1 \over \sqrt{3}}
\pmatrix{1 & 1 & 1 \cr 1 & \alpha & \bar \alpha \cr
1 & \bar \alpha & \alpha \cr}
\pmatrix{\chi_0 \cr \chi_1 \cr \chi_2 \cr}.
\eeq
It is also well-known that 
\beq
\eta^2(1/T) = \eta^2(T') = T \; \eta^2(T)
= -i \; e^{ F(T)} \; \eta^2(T).
\eeq
We define a polynomial $p$ which encodes the superpotential
couplings \myref{thetri}, up to a power of $\eta(T)$:
\beqa
p(T;\Phi_1,\Phi_2,\Phi_3) & = & \chi_0(T)
   (\Phi_1^1 \Phi_2^1 \Phi_3^1 + \Phi_1^2 \Phi_2^2 \Phi_3^2
   + \Phi_1^3 \Phi_2^3 \Phi_3^3) \nnn
& &   + \chi_1(T) (\Phi_1^1 \Phi_2^2 \Phi_3^3
   + \Phi_1^3 \Phi_2^1 \Phi_3^2 + \Phi_1^2 \Phi_2^3 \Phi_3^1 \nnn
& & \quad   + \Phi_1^3 \Phi_2^2 \Phi_3^1 + \Phi_1^1 \Phi_2^3 \Phi_3^2
   + \Phi_1^2 \Phi_2^1 \Phi_3^3).
\label{thepoly}
\eeqa
Using the transformation properties
enumerated above, it can be shown that
\beq
\eta^2 p \to -i \left( 1 \over iT \right) \eta^2 p
= -i e^{-F(T)} \eta^2 p.
\eeq
Thus, the functional $\eta^2 p$ transforms with
modular weight $-1$
up to a moduli independent phase $-i$, as required
by \myref{wcf}.  Here, we draw attention to the fact
that the monomials contained in \myref{thepoly}
do not by themselves transform in the required way.
Rather, it is the linear combination of fields in \myref{thepoly}
together with $\chi_i(T)$ factors which is modular
covariant, in the sense of \myref{wcf}.

Similar arguments hold for the axionic shift $T \to T' = T - i$.
Indeed, $\eta^2(T) \to \exp(i \pi /6) \eta^2(T)$
and \cite{chun:89,lauer:91a}
\beq
\pmatrix{\chi_0' \cr \chi_1' \cr \chi_2' \cr}
= e^{-i \pi /6}
\pmatrix{1 & 0 & 0 \cr 0 & \alpha & 0 \cr
0 & 0 & \alpha \cr}
\pmatrix{\chi_0 \cr \chi_1 \cr \chi_2 \cr},
\eeq
\beq
\pmatrix{{\Phi^1}' \cr {\Phi^2}' \cr {\Phi^3}' \cr}
= \pmatrix{\bar \alpha & 0 & 0 \cr 0 & 1 & 0 \cr
0 & 0 & 1 \cr}
\pmatrix{\Phi^1 \cr \Phi^2 \cr \Phi^3 \cr}.
\eeq
Then it can be checked that $\eta^2 p \to \eta^2 p$,
which transforms as it should (E.g., $c=0,d=1$ for the
axionic shift).

A general $SL(2,\Zbf)$ transformation \myref{sl2z} can
be built up out of the two operations analyzed above.
Thus, we are assured that the trilinear superpotential
coupling \myref{thetri} is modular covariant under
the entire duality group.
To summarize, for the general $SL(2,\Zbf)$ transformation
the polynomial \myref{thepoly} and the Dedekind 
$\eta$ function transform
as
\beqa
p(T;\Phi_1,\Phi_2,\Phi_3) & \to & e^{i \phi(a,b,c,d)} 
e^{-\sum_{n=1}^3 q_n F(T)}
p(T;\Phi_1,\Phi_2,\Phi_3), \nnn
\eta^2(T) & \to & e^{i \gamma(a,b,c,d)} e^{F(T)} \eta^2(T) , \nonumber
\eeqa
where $q_n$ is the modular weight of
the matter fields of species $n$, $\Phi_n^j$.  
It can then be checked that \myref{wcf}
holds for the trilinear superpotential 
term coupling \myref{thetri}.

Twisted couplings of dimension $3m >3$ in the effective
field theory remain to be calculated from the
underlying conformal field theory.  However, these
computations appear formidable.  As an example, we briefly
consider the form of six dimensional
twisted couplings using the
methods of \cite{dixon:87}.  The classical
action $S_{cl}$ may determined from {\it monodromy
conditions} on the underlying bosonic fields
$X(z,\bar z),\bar X(z,\bar z)$ of the 2-d orbifold,
where $z,\bar z$ provide a parameterization
of the string world-sheet; for each classical
solution $X_{cl}(z,\bar z)$, {\it local}
monodromy conditions demand
\beqa
\p_z X_{cl}(z) & = & {a(z_4,z_5,z_6) \over
[z(z-1)(z-z_4)(z-z_5)(z-z_6)]^{2/3}}, \nnn
\bar \p_{\bar z} X_{cl}(\bar z) &
= & {b(\bar z_4, \bar z_5, \bar z_6) \over
[\bar z(\bar z-1)(\bar z-\bar z_4)(\bar z-\bar z_5)(\bar z-\bar z_6)]^{1/3}}, \nnn
\bar \p_{\bar z} \bar X_{cl}(\bar z) &
= & {\tilde a(\bar z_4, \bar z_5, \bar z_6) \over
[\bar z(\bar z-1)(\bar z-\bar z_4)(\bar z-\bar z_5)(\bar z-\bar z_6)]^{2/3}}, \nnn
\p_z \bar X_{cl}(z) & = & {\tilde b(z_4,z_5,z_6) \over
[z(z-1)(z-z_4)(z-z_5)(z-z_6)]^{1/3}}.
\label{axc}
\eeqa
Here, $z_4,z_5,z_6$ are the three vertex insertion
locations which cannot be fixed by $SL(2,\Cbf)$
invariance while the first three vertices $z_1,z_2,z_3$
have been sent to $0,1,\infty$ resp.  
The functions $a,b,\tilde a,\tilde b$
depend on the unfixed vertex locations
and are determined for each classical solution $X_{cl}$
using {\it global} monodromy conditions.
The classical action is given by
\beq
S_{cl} = {1 \over 4\pi} \int d^2z \,
(\p X_{cl} \bar \p \bar X_{cl}
+ \bar \p X_{cl} \p \bar X_{cl}).
\label{scl}
\eeq
Upon substitution of formulae \myref{axc}
into \myref{scl}, one finds it necessary to
perform the integrals
\beq
I_r = \int {d^2z \over |
z(z-1)(z-z_4)(z-z_5)(z-z_6)|^r},
\qquad r={4 \over 3},{2 \over 3}.
\eeq
Using techniques developed in \cite{kawai:86a},
one can show that both of these integrals may
be written in the form of sums of products
of integrals along the real axis.  The
expressions obtained are typically
complicated special functions of the
unfixed vertex locations.  Rather than
attempting the calculation, we merely
write the results symbolically as\footnote{
At the risk of annoying the reader,
we have explicitly shown the dependence
on unfixed vertex locations, in order that
we might stress wherein the difficulty
lies.}
\beq
I_r = \sum_i c^i_r
F_i(r;z_4, z_5, z_6)
\bar G_i (r;\bar z_4, \bar z_5, \bar z_6).
\eeq
It follows that
\beq
S_{cl}(z_4, z_5, z_6; \bar z_4, \bar z_5, \bar z_6)
= {1 \over 4\pi}\left(a \tilde a I_{4/3} + b \tilde b I_{2/3}\right)
(z_4, z_5, z_6; \bar z_4, \bar z_5, \bar z_6).
\eeq
The functions
$a,b,\tilde a,\tilde b$ are also complicated
special functions of the unfixed vertex locations.
It should be clear that $S_{cl}$ is a horrendous function.
What is more, this action must be exponentiated, summed over an
infinity of classical solutions $X_{cl}$,
and multiplied by the quantum
part of the partition function to obtain the
correlator:
\beqa
& & \vev{ V_1(0,0) V_2(1,1) V_3(\infty,\infty)
V_4(z_4,\bar z_4) V_5(z_5,\bar z_5) V_6(z_6,\bar z_6) } \nnn
& & \qquad = Z(z_4, z_5, z_6; \bar z_4, \bar z_5, \bar z_6) \nnn
& & \qquad = Z_{qu}(z_4, z_5, z_6; \bar z_4, \bar z_5, \bar z_6)
\cdot \sum_{X_{cl}}
\exp [- S_{cl} (z_4, z_5, z_6; \bar z_4, \bar z_5, \bar z_6)] . \nnn
& &
\eeqa
Here, the quantum part of the partition function
$Z_{qu}(z_4, z_5, z_6; \bar z_4, \bar z_5, \bar z_6)$
will be some other horrific
function of the unfixed vertex locations.
Finally, we must extract the effective field theory
coupling coefficient by integrating the unfixed vertex locations
over the complex plane:
\beq
f_{i_1 \cdots i_6} \propto
\int \, d^2z_4 \, d^2z_5 \, d^2z_6 \,
Z(z_4, z_5, z_6; \bar z_4, \bar z_5, \bar z_6).
\label{cftcalc}
\eeq
Integrating over exponentials of sums
of products of special
functions of several complex variables is bad enough, but
one also has the $Z_{qu}$ prefactor
and the functions $a,b,\tilde a,\tilde b$
to deal with.  Suffice it to say, the explicit
calculation of coupling coefficients
for higher dimensional twisted couplings
certainly looks like a major undertaking.

As a result, we take a more phenomenlogical approach,
using modular covariance as a guide.  In this respect
our effective field theory is ``string-inspired'' rather than
``string-derived''.  It is our hope that by appealing
to symmetries of the underlying theory, we will capture
the most important features of the {\it bona fide} couplings.
Modular covariant $3m$-dimensional twisted couplings
can be constructed by tensoring the polynomials \myref{thepoly}.
This leads us to the implicit definition
of $T$-dependent twisted coupling
coefficients $f^T_{i_1 \cdots i_{3m}}$ given by
\beq
\sum_{ \{ i_j \}} f_{i_1 \cdots i_{3m}}^T
\Phi_1^{i_1} \cdots \Phi_{3m}^{i_{3m}} 
= {1 \over m! (3!)^m }
{\sum_{\{ n_j \}}}^{\prime} \prod_{k=0}^{m-1}
p(T;\Phi_{n_{3k+1}}, \Phi_{n_{3k+2}}, \Phi_{n_{3k+3}}),
\label{fdef}
\eeq
Here, $\sum_{\{ n_j \}}^{\prime}$ indicates
that the $3m$-tuple of subscripts
$(n_1, \ldots, n_{3m})$ should be summed
over all permutations of $(1,2,\ldots,3m)$.
The factor $1/m! (3!)^m$ accounts
for trivial permutation symmetries.
This construction treats the different species
of fields $\Phi_1, \ldots, \Phi_{3m}$
in a symmetric way with respect to
fixed point couplings.
If we define the $3m$-dimensional twisted
superpotential coupling as
\beq
\lambda \,
f_{i_1 \cdots i_{3m}}^T \, \,
[\eta(T)]^{2(\sum q_n -1)} \,
\Phi_1^{i_1} \cdots \Phi_{3m}^{i_{3m}},
\label{gtw}
\eeq
it can be checked that this will satisfy the requirement
\myref{wcf}.  We generically denote
the overall modulus- and fixed-point-independent 
coupling strength by $\lambda$.  The actual
value of this strength will depend on the dimensionality
of the coupling and the species of
fields which enter.  For $m>1$ in \myref{gtw}, it
is likely that $\lambda \gg 1$,
due to the numerous world-sheet
integrals which must be performed
in \myref{cftcalc} above, as pointed
out recently by Cveti\v c, Everett and Wang \cite{cvetic:98}.
This aspect of nonrenormalizable couplings
in string-derived models has been overlooked in much of the
earlier literature, due to the temptation
to estimate unknown quantities as $\ordnt{1}$.

All of the above considerations dealt with a two-dimensional
orbifold.  We must generalize
our results to the six-dimensional case.
Also, it is necessary to say what should be
done if untwisted states appear in a coupling or
if some of the twisted states have nonzero oscillator
numbers.  Here, we address these complications
only to the extent that it is necessary for the
quark mass couplings in the toy model 
which is to be discussed in Section~\ref{pckm}.  Again, we
take a phenomenological approach, using modular
covariance as a guide, rather than attempting to
explicitly derive effective field theory coupling
coefficients from the underlying
conformal field theory.  The reason once
again is that the calculations appear to
be extremely difficult.

For a six-dimensional $Z_3$ orbifold the twist field
operators generalize to $\s_{ijk}(z,\bar z)$,
where the triple $ijk$ denotes which of the
27 fixed points the twisted field $\Phi_n^{i j k}$
created by the vertex operator sits at.
The indices refer to the fixed point
locations in each of the three complex planes.
To the extent that the three complex planes
are orthogonal to each other, which is the
case if the off-diagonal T-moduli vanish,
it is possible to
decompose the twist operators into
a tensor product of two-dimensional orbifold
twist operators:
\beq
\s_{ijk} (z, \bar z) \equiv\s_i^1 (z, \bar z)
\otimes \s_j^2 (z, \bar z) \otimes \s_k^3 (z, \bar z).
\eeq
Then the six-dimensional orbifold couplings
are a tensor product of the two-dimensional
ones.  Indeed, it can be checked that
given the transformation properties of the
two dimensional twist operators,\footnote{Here,
$\tilde \s_i^I = e^{i\beta(T^I,\bar T^{\bar I})/2} \s_i^I$
generalizes the $\tilde \s_i$ defined above
for the two-dimensional orbifold case.}
\beqa
& & \prod_{I=1}^3 p(T^I; \tilde \s^I (z_1, \bar z_1) \, \tilde \s^I (z_2, \bar z_2)
\, \tilde \s^I (z_3, \bar z_3)) \to \nnn
& & \qquad \prod_{I=1}^3 e^{i \phi(a^I,b^I,c^I,d^I)}
p(T^I; \tilde \s^I (z_1, \bar z_1) \, \tilde \s^I (z_2, \bar z_2)
\, \tilde \s^I (z_3, \bar z_3)).
\eeqa
This has the correct transformation, up to the
factor $\prod_I \exp(-\sum_i q_i^I F(T^I))$ which would come
from the transformation properties of the
matter field vertex operators not accounted for by $\tilde \s_i^I$.
Then the generalization of \myref{thetri} with the
correct modular transformation properties is:
\beq
\lambda \cdot \left(
\prod_I [\eta(T^I)]^2 \, f_{i_1^I i_2^I i_3^I}^{T^I}
\right) \,
\, \Phi_1^{i_1^1 i_1^2 i_1^3}
\Phi_2^{i_2^1 i_2^2 i_2^3} \Phi_3^{i_3^1 i_3^2 i_3^3}
\label{gentri}
\eeq
It is easy to check that this holds for the higher
dimensional couplings as well.  We simply take
the obvious products of 2-d orbifold coupling
coefficients:
\beq
\lambda \cdot \left(
\prod_I [\eta(T^I)]^{2(\sum_n q_n^I - 1)}
f_{i_1^I \cdots i_{3m}^I}^{T^I} \right)
\Phi_1^{i_1^1 i_1^2 i_1^3} \cdots
\Phi_3^{i_{3m}^1 i_{3m}^2 i_{3m}^3} .
\label{gchd}
\eeq

Now we consider the occurence of twisted
oscillator fields and untwisted fields in
higher dimensional couplings.  The vertex operator
for a twisted oscillator field is proportional
to an excited twist operator.  The excited
twist operator can be written
in terms of an ordinary twist operator and
a factor of $\dxb{\ell}$, with $\ell$ depending on the
oscillator direction \cite{dixon:87}.
Then the classical action $S_{cl}$ is still
computed in the presence of the same
twist operators and we expect that the fixed
point dependence should be the same as
in the case where none of the twisted
states were oscillators.  Indeed,
this has been found to be the case in $Z_N$
orbifolds where renormalizable couplings
may involve oscillator fields \cite{baili:94a}.
It has been shown \cite{ibane:92,blmw}
that the modular weight of an $N_L = 1/3$ oscillator state
$Y^{\ell, ijk}$ (where $\ell$ is the direction of the
oscillator and $ijk$ specifies the fixed
point location) is given by
\beq
q^I(Y^{\ell, ijk}) = (2/3,2/3,2/3) + \delta^I_\ell.
\eeq
Since the vertex operator creating the twisted
oscillator state is proportional to $\tilde \s_i^1 
\otimes \tilde \s_j^2 \otimes \tilde \s_k^3$, we expect that the 
state transforms under $T^I \to 1/T^I$
according to \myref{twtrd} if $I \not= \ell$, while
the power $2/3$ should be replaced by $5/3$ if $I=\ell$.
Based on this assumption, the modular invariant
couplings involving oscillator fields (which 
for the $Z_3$ orbifold are
always higher dimensional couplings because of the
automorphism selection rule) are obtained from
\myref{gchd} directly, with the oscillator nature
of states reflected in the modular weights $q_i^I$ and
a different overall strength $\lambda$ than would
be obtained if the states were not oscillators,
due to the presence of the additional operator $\dxb{\ell}$.
Obviously, adding untwisted states to a coupling does not introduce
any new twist operators.  We can always choose a
``picture'' such that the vertex operator of the
untwisted state $U^i$ goes like $\dx{i}$.  Then the
change in the conformal field theory correlator will
be completely in the quantum part and we expect the
fixed point dependence to be unchanged.  An untwisted
state $U^i$ has modular weight $q_i^I = \delta_i^I$.
As a result of these arguments, we conclude that the
coefficients for a coupling with $3m$ twisted
fields and $n$ untwisted fields can be read off from
\myref{gchd}, only we must include the modular
weights of the untwisted fields in the sums in the exponents
of the $\eta$ functions, and the overall coupling
strength $\lambda$ will be different than if
no untwisted fields were in the coupling.

%
%

\mysection{Toy model}
\label{pckm}
Below the scale of \ux\ breaking, $\Lambda_X \sim 10^{17}$ GeV,
the quark and higgs spectrum
is assumed to be that of the Minimal Supersymmetric
Standard Model (MSSM).  Extra color
triplets get vector mass couplings when Xiggses
get vevs.  As a consequence,
they get masses $\ordnt{\Lambda_X}$
and integrate out of the spectrum
near the string scale.
Above $\Lambda_X$ we suppose
the spectrum contains untwisted doublet quarks $Q^i$,
untwisted $u_L^c$-like quarks $u_1^i$,
twisted $u_L^c$-like quarks $u_2^i$,
twisted $d_L^c$-like quarks $d^i$,
untwisted $H_u$-like higgs doublets $H_u^i$
and twisted $H_d$-like higgs doublets $H_d^i$.
These assignments are quite similar to those of the
FIQS model.  The superscript on untwisted fields
corresponds to the H-momentum of the states
in the underlying theory, and takes values $i=1,2,3$.
The superscript on twisted fields corresponds
to the fixed point location in the third complex plane
of the six-dimensional compact space.  As in the FIQS model, three
linear combinations of the six $u_L^c$-like
quarks survive in the low-energy spectrum,
which we describe by mixing matrices $X^1$ and $X^2$:
\beq
u_1^i = X^1_{ij} u_{jL}^c + \mtxt{heavy},
\qquad
u_2^i = X^2_{ij} u_{jL}^c + \mtxt{heavy}.
\label{umx}
\eeq
The mixings to $\Lambda_X$ scale mass eigenstates,
denoted by ``heavy'', are not important to
our tree level analysis of low energy quark
mass matrices.
We assume that all extra higgses in\-te\-grate
out near the scale $\Lambda_X$ due to vector
couplings induced by the Xiggs vevs 
(as in the FIQS model), leaving one pair
which we identify as the $H_u$ and $H_d$ of the MSSM:
\beq
H_u = H_u^1, \qquad H_d = H_d^3.
\label{hid}
\eeq
We introduce SM singlet Xiggses $Y_1^{\ell i},Y_2^{\ell i}$ and 
$Y_3^{\ell i}$ which get $\ordnt{\Lambda_X}$ vevs
and appear in the nonrenormalizable mass couplings of
the quarks.  The $Y$'s are charged under \ux\
and their scalar components are among those
which get vevs to cancel the \ux\ FI term.
The $Y$'s are twisted states with nonzero left-moving oscillator
number $N_L = 1/3$.  The first superscript corresponds to the
oscillator direction in the compact space, $\ell = 1,2,3$.
The second superscript corresponds to the fixed
point location in the third complex plane of the
compact space.  Such fields also arise in the FIQS model.
In what follows, we use the same symbols for superfields
and scalar components of fields other than the quarks, with
the meaning obvious by context.  Similarly, whether we refer
to a quark superfield or its fermionic component should
be obvious by context.

We assume that the leading holomorphic couplings
giving quarks masses are contained in the superpotential
\beqa
W_{qm} & = & \lambda_0 |\e_{ijk}| H_u^i Q^j u_1^k  
+ \lambda_1 \lambda_{i_1 i_2 i_3 \ell_1 \ell_2 k j}
Y_1^{\ell_1 i_1} Y_1^{\ell_2 i_2} H_u^k Q^j u_2^{i_3}
\nnn
& & + \lambda_2 
\lambda_{i_1 i_2 i_3 i_4 i_5 i_6 \ell_1 \ell_2 \ell_3 \ell_4 j}
Y_1^{\ell_1 i_1} Y_2^{\ell_2 i_2} 
Y_3^{\ell_3 i_3} Y_3^{\ell_4 i_4}
H_d^{i_5} Q^j d^{i_6}.
\label{psup}
\eeqa
The trilinear untwisted coupling is proportional to $|\e_{ijk}|$
according to the {\it conservation of H-momentum} orbifold
selection rule.
The fields $Y_{1,2,3}^{\ell i}$ are assumed to
be charged identically under \ux.
We assume that the 
operators contained in \myref{psup}
are each neutral under the full rank sixteen gauge group
$G$ obtained from the orbifold embedding.
Thus, the other fields are assumed to have \ux\ charges\footnote{
In the three generation constructions presently
under consideration, the \ux\ charges are
independent of generation number.
Thus, horizontal flavor symmetries,
such as those considered in refs.~\cite{farag:94a,dudas:96a},
cannot be implemented in this context.}
such that each coupling is \ux\ neutral.
We assume that off-diagonal
T-moduli $T^{I \bar J}, \; I \not= \bar J,$ 
have vanishing vevs, as in the FIQS model, so that
the leading order kinetic terms for the matter fields are
diagonal.  In the FIQS model, nonvanishing
off-diagonal T-moduli 
lead to nonvanishing F-terms
which break supersymmetry at the scale $\Lambda_X$,
which is unacceptable.  Though these fields may acquire
vevs once supersymmetry is broken in the hidden sector,
we expect the vevs to be at most of order the hidden
sector supersymmetry breaking scale.  As a result,
off-diagonal T-moduli give negligible contributions
to the kinetic terms of the quarks.
We further assume that the diagonal T-moduli
$T^I \equiv T^{I \bar I }$ are stabilized at
one of their self-dual points $\vev{T^I} = 1,e^{i\pi/6}$
once supersymmetry is broken,
consistent with models of hidden sector
supersymmetry breaking by gaugino condensation \cite{binet:97a,bkl}.
It has also been argued that the T-moduli
may stabilize at other points on the
unit circle \cite{bkl}.  Either way, it would
appear that string-derived scalar potentials for the T-moduli
stabilize them to values $|T^I|=1$.
For this reason we view models which allow
$\vev{T^I}$ as large as required to obtain
hierarchies in the Yukawa couplings of twisted
fields \cite{lth} to be unmotivated.

The K\"ahler metric for matter
fields in (0,2) $Z_3$ orbifolds 
(arbitrary Wilson lines and point
group embeddings)
has been determined to leading
order \cite{blmw,baili:92a}.
In the case of
vanishing off-diagonal
T-moduli, the metric of 
the untwisted fields $Q^i$ and $u_1^i$ is given by
\beq
\vev{K_{Q}}_{i \bar \ell}= \vev{K_{u_1}}_{i \bar \ell}
= \delta_{i \bar \ell}
\vev{T^i + \bar T^\bari}^{-1}.
\eeq
We make redefinitions $Q^i \to 
\vev{T^i + \bar T^{\bari}}^{1/2}
Q^i$ and similarly for $u_1^i$.
Similar arguments hold for the twisted fields $u_2^i$
and $d^i$, whose K\"ahler metric at leading order is
\beq
\vev{K_{d}}_{i \bar \ell} =
\vev{K_{u_2}}_{i \bar \ell} = \delta_{i \bar \ell}
\prod_{I} \vev{T^I + \bar T^{\bar I}}^{-2/3} .
\eeq
We assume that the overall coupling strengths
$\lambda_0, \lambda_1, \lambda_2$ in
\myref{psup} reflect these rescalings
and that the quark fields
entering these couplings are the rescaled
ones.  We also assume that the
factor $\exp \vev{K}/2$ has been absorbed
into these coupling strengths and make use of
the fact that 
$\vev{{\mathrsfs D}_i D_j W} \approx \vev{W_{ij}}$
is a very good approximation,
in the supergravity lagrangian notation of
Wess and Bagger \cite{wb}.  It can be checked
that the terms which we drop are
$\ordnt{m_{\tilde G} m_W / m_P}$,
where $m_{\tilde G}$ is the gravitino mass.
When working with these rescaled
fields, we may raise and lower indices
with impunity since their K\"ahler metric
is canonical in the leading order approximation
made here.
Once the fields $u_1^i$ and $u_2^i$ have been
rescaled in this way, the mixings to mass
eigenstates $u_{iL}^c$ and their three heavy
relatives (all canonically normalized)
can be made unitary.  We then have as a
constraint:
\beq
(X^{1\dagger} X^1)_{jk} + (X^{2\dagger} X^2)_{jk}
= \delta_{jk} .
\label{uct}
\eeq
When one takes \myref{umx} and \myref{hid} into account,
the effective Yukawa couplings for the quarks are given by
\beqa
\lambda_{jm}^u & = &
\lambda_0 (\delta_{j2} X^1_{3m} + \delta_{j3} X^1_{2m})
+ \lambda_1 \lambda_{i_1 i_2 i_3 \ell_1 \ell_2 1 j}
\vev{ Y_1^{\ell_1 i_1} Y_1^{\ell_2 i_2} } X^2_{i_3 m} ,
\nnn
\lambda_{jm}^d & = &
\lambda _2 \lambda_{i_1 i_2 i_3 i_4 3 m \ell_1 \ell_2 \ell_3 \ell_4 j}
\vev{ Y_1^{\ell_1 i_1} Y_2^{\ell_2 i_2} 
Y_3^{\ell_3 i_3} Y_3^{\ell_4 i_4}} .
\label{dcp}
\eeqa
In going from \myref{psup} to 
\myref{dcp}, we have set $k=1$ in the second
coupling of \myref{psup} because $H_u = H_u^1$
and we have fixed $i_5 = 3$ and $i_6 = m$ in the
third coupling of \myref{psup} since
we couple to $H_d = H_d^3$ and $d^m$.

We now apply the assumptions of Section \ref{modinv} to
the toy model.  The toy model
is based on string-derived models where two ``discrete'' Wilson lines
are included in the embedding to give a three
generation model \cite{font:90a,thrgen}.  
By construction, states which
differ only by their fixed point location in the
third complex plane have identical gauge quantum
numbers.  On the other hand, states which differ
by fixed point locations in the first two complex
planes generally have different gauge quantum
numbers under the rank 16 gauge group which 
survives the orbifold compactification.  
Typically, the embedding is arranged so that the
rank 16 gauge group has the form $SU(3) \times
SU(2) \times [U(1)]^m \times G_c$, where $G_c$ is
a simple group which condenses in the hidden sector
to break supersymmetry.  The extra $U(1)$'s get
broken down to $U(1)_Y \times$ (hidden $U(1)$'s) 
by the FI term associated
with the anomalous \ux.
Fixed point locations in the first two complex
planes become species labels.
In what follows, the only fixed point location superscript
on twisted states is that corresponding to the
third complex plane.  This serves as a family number
for twisted states in these models.  Twisted ``relatives''
differ only by their fixed point location in the
third complex plane, so in many respects the
effective Yukawas behave as if we were working
with a two-dimensional orbifold.  
The coupling coefficients in
\myref{dcp} are given by:
\beqa
\lambda_{i_1 i_2 i_3 \ell_1 \ell_2 1 j}
& = &
\left( \prod_I \eta(T^I)^{2(Q_1^I - 1)} \right)
\chi_0(T^1) \chi_0(T^2)
f_{i_1 i_2 i_3}^{T^3} \nnn
& & \qquad \qquad \mtxt{if} \quad (\ell_1,\ell_2)
= (\underline{1,j}), \nnn
& = & 0 \qquad \mtxt{else}; \\
\lambda_{i_1 i_2 i_3 i_4 3 m \ell_1 \ell_2 \ell_3 \ell_4 j}
& = &
\left( \prod_I \eta(T^I)^{2(Q_2^I - 1)} \right)
\chi_1(T^1)^2 \chi_1(T^2)^2
f_{i_1 i_2 i_3 i_4 3 m}^{T^3} \nnn
& & \qquad \qquad \mtxt{if} \quad (\ell_1,\ell_2,\ell_3,\ell_4)
= (\underline{1,2,3,j}), \nnn
& = & 0 \qquad \mtxt{else}; \label{fpn}
\eeqa
\beq
Q_1^I = 2 + 2 \delta_1^I + 2 \delta_j^I, 
\qquad Q_2^I = 5 + 2 \delta_j^I.
\eeq
The constraints on the allowed values of $\ell_i$
follow from the automorphism selection rule;
underlining denotes that any permutation of entries
is permitted.  The coefficients 
$f_{i_1 i_2 i_3}^T$ carry the dependence on
third complex plane fixed point locations
of twisted fields appearing in the nonrenormalizable
up-type quark mass coupling, and are
given explicitly in \myref{dff} above.
The third complex plane fixed point
dependence for the down-type quark mass coupling
follows from the six-dimensional twisted coupling,
and is defined implicitly by \myref{fdef}.
The six twist coupling 
coefficients $f_{i_1 i_2 i_3 i_4 i_5 i_6}^T$ vanish
by the {\it lattice group selection rule} unless
$ i_1 + \cdots + i_6 = 0 \mod 3$.
It can be checked that the choices $i_1,
\ldots, i_6$ satisfying this rule can be
divided into four classes, depending on whether
triples of the indices can be formed where
the entries of the triples are either all
the same ({\it s}) or all different ({\it d}).  Members
of the same class have identical values for
$f_{i_1 i_2 i_3 i_4 i_5 i_6}^T$.  The nonvanishing
values of $f_{i_1 i_2 i_3 i_4 i_5 i_6}^T$
are given in Table \ref{mytable6}, according to
which of the four classes the indices belong
to.  A representative example $(i_1 \cdots i_6)$
for each class is given to avoid any confusion.
The factor of $\chi_1(T^1)^2 \chi_1(T^2)^2$
in \myref{fpn}
follows from an additional assumption of
our model:  the fixed point locations
(of the six species of twisted fields in the down-type Yukawa
coupling) in the first two complex planes
are such that the lattice group
selection rule in each of the two planes
is satisfied in the ({\it dd}) way of Table \ref{mytable6}.
\begin{table}
\begin{center}
$$
\begin{array}{|c|c|c||c|c|c|} \hline
\mtxt{class} & f_{i_1 \cdots i_6}^T &
\mtxt{representative} &
\mtxt{class} & f_{i_1 \cdots i_6}^T &
\mtxt{representative} \\ \hline \hline
ss & 10 \chi_0(T)^2 & (111111) &
sd & 2 \chi_0(T) \chi_1(T) & (111123) \\ \hline
ss' &  \chi_0(T)^2 & (111222) &
dd &  \chi_1(T)^2 & (123123) \\ \hline
\end{array}
$$
\end{center}
\caption{Six twist coupling coefficients by class}
\label{mytable6}
\end{table}

The strengths $\lambda_0, \lambda_1, \lambda_2$,
the mixing matrices $X^1_{ij}$,
$X^2_{ij}$, and the background $\vev{Y_{1,2,3}^{\ell i}}$
are treated as phenomenological parameters.
We tune the couplings, mixings and
vevs to values which yield a reasonable
phenomenology.  In principle, all of these
quantities would be fixed by a full and complete
analysis of the string-derived effective
supergravity.  However, some of
the background fields in $\vev{Y_{1,2,3}^{\ell i}}$
are D-moduli \cite{gaill:00a}; in order to
fix these we must say how the D-moduli flat
directions are lifted.  In the reference just cited
it was suggested how these flat directions 
may be lifted by nonpertubative
effects in the hidden sector, via superpotential
couplings of the D-moduli to hidden sector
matter condensates.
It should also be noted that the 
symmetries which give rise to the D-moduli
are only valid for the classical
scalar potential under the assumption of
vanishing of F-terms for the
D-moduli in the background.  As a result,
they are pseudo-Goldstone bosons and
we expect that the D-moduli
flat directions will also be lifted by loop
corrections.\footnote{We thank Korkut Bardakci for
bringing this to our attention.}  Preliminary
estimates show that the D-moduli get contributions
to their masses of order the gravitino mass
from either effect.
We are curently exploring
how the phases of the Xiggses may be
fixed by these mechanisms and
what effect this will have on the
KM phase in models of the type discussed here.  Our results
will be presented elsewhere.

We do not intend to be exhaustive
in our analysis of the phenomenology
of \myref{dcp}.
Rather, we would simply like to
demonstrate that it is possible to obtain
a quark phenomenology which is consistent
with experimental data.  The shortest route
to this goal is to implement textures
in the effective Yukawa couplings.
In this way our scan over parameter space is biased toward
viable models.  We make use of the results of
a recent analysis of viable mass
matrices \cite{kuoma:99a}, though we will
not impose the hermiticity constraint implemented
there since we have no motivation for it
in the present context.  We impose
the textures
\beq
\lambda^u  = 
f_t \pmatrix{ A_u \cab^{8} & 0 & C_u \cab^4 \cr
          0 & D_u \cab^4 & E_u \cab^2 \cr
          \tilde C_u \cab^4 & \tilde E_u \cab^2 & 1 \cr}, \qquad
\lambda^d  = 
f_b \pmatrix{ 0 & B_d \cab^3 & 0 \cr
          \tilde B_d \cab^3 & D_d \cab^2 & 0 \cr
          0 & 0 & 1 \cr},
\label{ytx}
\eeq
through an arrangement of the mixing matrices
$X^1_{ij}, X^2_{ij}$ and vevs $\vev{Y_{1,2,3}^{\ell i}}$.
Here,
\beq
\cab \approx V_{us} \approx 0.22, \qquad
f_t \approx  m_t / \vev{H_u^0}, \qquad
f_b \approx m_b / \vev{H_d^0},
\eeq
\beq
A_u, C_u, \tilde C_u, D_u, E_u, \tilde E_u, B_d,
\tilde B_d, D_d \sim \ordnt{1}.
\label{ooc}
\eeq
The forms \myref{ytx} were obtained
by imposing the following textures in
$X^{1,2}$ and $Y_{1,2,3}$:
\beq
X^1 =
\pmatrix{ * & * & * \cr
          r_1 \cab^4 & r_2 \cab^2 & r_3 \cr
          0 & r_4 \cab^4 & r_5 \cab^2 \cr}, \qquad
X^2 =
\pmatrix{ * & * & * \cr
          * & * & * \cr
          s_1 \cab^4 & 0 & s_2 \cr},
\label{x12a}
\eeq
\beq
\vev{Y_1^{\ell i}} = y_1 \cab^2 \delta^\ell_1 \delta^i_3, \qquad
\vev{Y_2^{\ell i}} = y_2 \delta^\ell_2 \delta^i_3, \qquad
\vev{Y_3} = \pmatrix{ y_{31} \cab^3 & 0 & 0 \cr
          y_{32}\cab^2 & \tilde y_{31} \cab^3 & 0 \cr
          0 & 0 & y_{33} \cr},
\eeq
The elements denoted by $*$ in \myref{x12a} are left
unspecified since they do not appear in the effective
Yukawa matrices.  We assume that they are chosen
such that \myref{uct} is satisfied, which
is generally true provided
$|r_3|^2+|s_2|^2+|r_5|^2 \cab^4 \leq 1$
because of the $\cab^n$ suppressions on entries
of the other columns.
Up to this restriction, the quantities
$r_i, s_i, y_i, y_{ij}, \tilde y_{31} \sim \ordnt{1}$.
We note that there is no inconsistency in having
Xiggs vevs larger than the FI term, since the vevs
of fields having opposite \ux\ charge can be played
off against each other in the \ux\ D-term.  As an example,
in the FIQS model the Y-type Xiggses can be made
arbitrarily large while maintaining D-flatness
by simultaneously increasing some of the
vevs of non-oscillator Xiggses which they denote
by $S_6^i$.  Of
course at some point the nonlinear $\s$-model
perturbation theory breaks down.

Given the assumptions enumerated above, the
effective Yukawa matrices take the form
($T^I_c \equiv \vev{T^I}$):
\beq
\lambda^u = \lambda_0 \pmatrix{ h_u s_1 \cab^8 & 0 & h_u s_2 \cab^4 \cr
   0 & r_4 \cab^4 & r_5 \cab^2 \cr
   r_1 \cab^4 & r_2 \cab^2 & r_3 },
\label{eyu}
\eeq
\beq
h_u \equiv {\lambda_1 \over \lambda_0} \eta(T_c^1)^{10}
\eta(T_c^2)^2 \eta(T_c^3)^2 \chi_0(T_c^1) \chi_0(T_c^2)
\chi_0(T_c^3) y_1^2,
\eeq
\beq
\lambda^d = h_d \pmatrix{0
   & 2 \eta(T_c^1)^4 \chi_1(T_c^3) y_{31} \cab^3
   & 0 \cr
   2 \eta(T_c^2)^4 \chi_1(T_c^3) \tilde y_{31} \cab^3
   & 2 \eta(T_c^2)^4 \chi_1(T_c^3) y_{32} \cab^2
   & 0 \cr
   0 & 0 & 5 \eta(T_c^3)^4 \chi_0(T_c^3) y_{33} \cr },
\label{eyd}
\eeq
\beq
h_d \equiv 2 \lambda_2 \left[ \eta(T_c^1)
\eta(T_c^2) \eta(T_c^3) \right]^8
\chi_1(T_c^1)^2 \chi_0(T_c^2)^2 \chi_0(T_c^3) y_1 y_2 y_{33} \cab^2 .
\eeq
The quantities $B_d, \tilde B_d, D_d$ in
\myref{ytx} can be varied independently
by adjusting the ratios $y_{31}/y_{33}, \tilde y_{31}/y_{33},
y_{32}/y_{33}$.  The ratio of heavy generation
Yukawa eigenvalues $f_b/f_t$
can be varied independently of $B_d, \tilde B_d, D_d$
and $\lambda^d$ by adjusting $y_2 y_{33}^2/y_1$.
The top
quark Yukawa eigenvalue $f_t$ can be adjusted independently
by varying $r_3$.  However, if $\lambda_0$ is too small
there may be a minimum $\tan \beta$ below which
we cannot match experimental data, since $|r_3|<1$
is required by \myref{uct}.  Recall that we have
absorbed a factor $\exp \vev{K}/2$ into $\lambda_0$,
as well as the effects of quark field rescalings
to account for noncanonical kinetic terms.  Typically,
$\exp \vev{K}/2 < 1$ when the string moduli get
$\ordnt{1}$ vevs, so this may be a worry.
Without an explicit model of the superpotential
couplings and Xiggs vevs which determine the
mixing matrices $X^{1,2}$,
it is not possible to say whether
or not the entries of $\lambda_u$ can be varied
independently of each other and $\lambda_d$;
we will assume that this is true.

With the above assumptions, scanning over
the Xiggs vevs and the mixing coefficients $r_i,s_i$
for viable models
is equivalent to varying the 
coefficients in \myref{ooc} independently
and tuning the values
of $f_t,f_b$ to agree with experimental data.
We then rephase the
quarks according to the convention
\beq
V_{ud} > 0, \quad V_{us} > 0, \quad V_{cb} > 0, \quad
V_{ts} < 0, \quad V_{cd} < 0,
\eeq
to which the Wolfenstein parameterization 
\cite{wolfe:84a,chaul:84a} is
an approximation.  As is well known, the advantage
of such a parameterization is that the elements with
significant complex phase are the smallest ones,
$V_{ub}$ and $V_{td}$.  

All of these calculations
are done at the \ux\ breaking scale, and are therefore
subject to evolution under the renormalization
group.  The evolution of the quark masses and
mixing angles assuming the MSSM spectrum has
been studied extensively; approximate analytic
formulas are available, for example in refs.~\cite{qrge,kx}.
We will use the approximations of \cite{kx}
to evolve the low energy data to the scale
of \ux\ breaking, which we assume to be
$\Lambda_X \sim \Lambda_s \sim 5 \times
10^{17}$ GeV, based on what occurs in the FIQS
model.\footnote{See Appendix \ref{normsub}.}
The following quantities are approximately
scale-independent:
\beq
{m_c \over m_u}, \quad {m_s \over m_d}, \quad |V_{ud}|,
\quad |V_{cs}|,\quad |V_{tb}|,\quad |V_{us}|,\quad |V_{cd}|.
\label{rg1}
\eeq
The running of the other quantities is approximately
given by:
\beq
\left. {m_t \over m_c} \right|_{\Lambda_X}
 = {1 \over \xi_t^3 \xi_b} \left. {m_t \over m_c} \right|_{M_Z},
\quad \left. {m_b \over m_s} \right|_{\Lambda_X}
 = {1 \over \xi_t \xi_b^3} \left. {m_b \over m_s} \right|_{M_Z},
\quad |V_{cb}|_{\Lambda_X} = \xi_t \xi_b |V_{cb}|_{M_Z};
\label{rg2}
\eeq
the quantities $|V_{ub}|,|V_{td}|,|V_{ts}|$
scale in the same manner as $|V_{cb}|$.
The scaling functions are given by
\beq
\xi_{t,b} = \exp \left[ {-1 \over 16 \pi^2}
\int_0^{\ln ( \Lambda_X / M_Z)}
d\chi \; f^2_{t,b}(\chi) \right],
\eeq
where $\chi=\ln ( \mu/M_Z)$ and
$f_{t,b}(\mu)$ are the Yukawa coupling
eigenvalues of the top and bottom quarks
appearing in \myref{ytx}, at the scale $\mu$.
We assume the scale of observable sector
supersymmetry breaking is 1 TeV and we
set $\tan \beta = 5$.  For low energy
data we use the values of the running
quark masses at the scale $M_Z$
as determined in ref.~\cite{fk} and
the CKM data listed in ref.~\cite{pdg}.  Taking
into account the errors quoted in these
two source, we find the following values:
$$
f_t(\Lambda_X) = 0.74^{+1.65}_{-0.24},
\qquad f_b(\Lambda_X) = 0.028(4),
$$
$$
\left. {m_t \over m_c} \right|_{\Lambda_X}
= 440^{+390}_{-100}, \qquad
\left. {m_c \over m_u} \right|_{\Lambda_X}
= 290(60),
$$
$$
\left. {m_b \over m_s} \right|_{\Lambda_X}
= 38(7), \qquad
\left. {m_s \over m_d} \right|_{\Lambda_X}
= 20(4),
$$
\beq
|V_{CKM}|_{\Lambda_X} =
\pmatrix{ 0.9752(8) & 0.220(4) & 0.0027(12) \cr
   0.220(4) & 0.9745(8) & 0.033(4) \cr
   0.014(11) & 0.065(36) & 0.9992(2) \cr }.
\label{hidta}
\eeq
We stress that theoretical errors due to
the approximations made in \myref{rg1}
and \myref{rg2} have not been included
in the estimates of uncertainty.  However,
for our purposes this is not an important
issue since we can always make a small shift
in the $\ordnt{1}$ parameters of our toy
model to account for small corrections and
larger uncertainties will just mean that more
points in parameter space will give viable
models.
\begin{figure}
\begin{center}
\includegraphics[height=5in,width=3.5in,angle=-90]{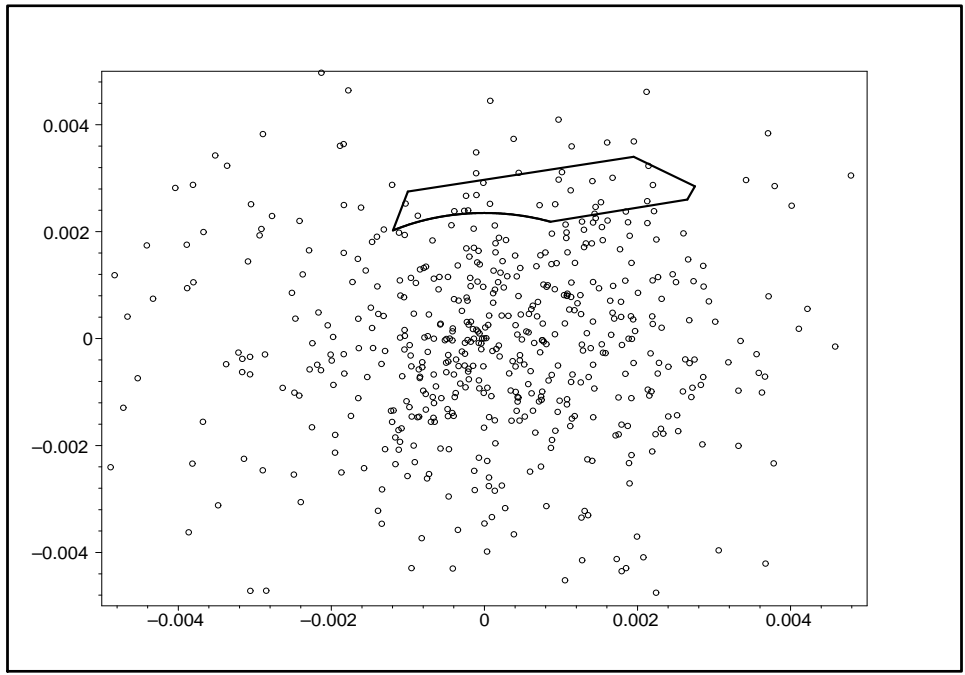}
\end{center}
\caption{$V_{ub}(M_Z)$ for complex mixing matrices
$X^{1,2}$ and Xiggs vevs $\vev{Y_{1,2,3}^{\ell i}}$.
The T-moduli are stabilized at $T_c^I=1$. 
For comparison, the experimentally preferred
region \cite{pdg} is outlined.}
\label{fig1}
\end{figure}

In our analysis we consider both generic
and extreme possibilities in order to get
a feel for how the KM phase depends on the
various sources of phases in \myref{eyu}
and \myref{eyd}.

\vspace{10pt}
\noindent {\it Case 1:  generic mixings and Xiggs vevs}
\vspace{10pt}

We have scanned over the magnitudes of the
parameters in \myref{ooc} with Gaussian
distributions centered on values
suggested by the central
values in \myref{hidta} and with
spreads suggested by the estimated
uncertainties.  Phases have been scanned on a flat
distribution over the interval $(-\pi,\pi]$.
We then compared mass ratios and the magnitudes
of CKM elements, except $|V_{ub}|$,
to the values in \myref{hidta}; if these
results agreed with the values in \myref{hidta}, 
except $|V_{ub}|$,
up to the stated uncertainties, we stored
the values of $V_{ub}(\Lambda_X)$.  We
then scaled the magnitude of $V_{ub}$
according to \myref{rg2} but left the phase
unrotated to get an estimate of $V_{ub}(M_Z)$.
In Figure \ref{fig1}
we plot our results, showing only points
near the acceptable
region.  The results are hardly surprising:
if we allow the phases of the fields Xiggses
to float randomly, the KM phase can take on any value
we like.  No magical cancellation occurs.
Although regions of parameter space in this
toy model do
exist which have reasonable quark masses,
mixings and CP violation, the
model provides no understanding of why we live
in one region of parameter space rather than
another.  All that can be said is that our model,
which contains many more free parameters than
the number of experimental data points which
we are attempting to fit, can be made to agree
with what is known about the quark sector.
One promising point does emerge, however.
Figure \ref{fig1} shows that CP violation is
generic in the toy model under consideration.
To be fair, one could argue that we have gone
through a lot of unnecessary work to prove the obvious:
if nonrenormalizable couplings contribute
significantly to the effective quark Yukawa
matrices, and the Xiggses in these
nonrenormalizable couplings get complex vevs,
then CP violation is to be expected.
However, as we discussed in Section \ref{intro},
one can wonder whether the symmetry constraints
of modular invariance and orbifold
selection rules might
render these phases spurious.  We have
explicitly shown that this is not the case.

\vspace{10pt}
\noindent {\it Case 2:  complex Xiggs vevs}
\vspace{10pt}

\begin{figure}
\begin{center}
\includegraphics[height=5in,width=3.5in,angle=-90]{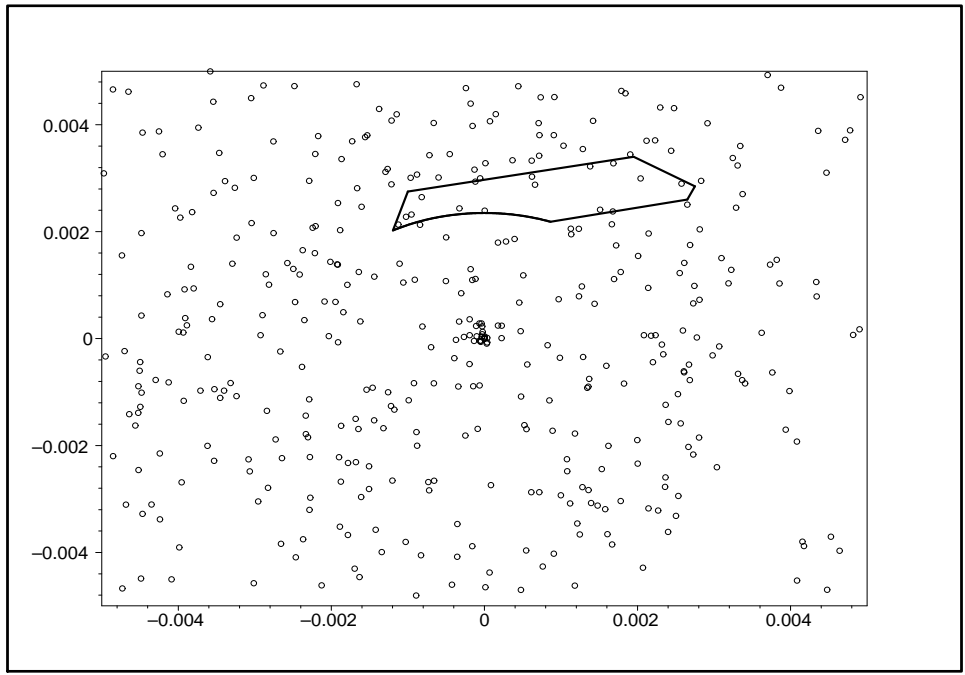}
\end{center}
\caption{$V_{ub}(M_Z)$ for real mixing matrices
$X^{1,2}$ and complex Xiggs vevs $\vev{Y_{1,2,3}^{\ell i}}$.
The T-moduli are stabilized at $T_c^I=1$.}
\label{fig3}
\end{figure}

Here, we make the quantities $r_i,s_i$ in 
\myref{x12a} real and positive and keep $T_c^I=1$
in order to isolate the effects of the
phases of the Xiggs.  With these assumptions
it can be seen from \myref{eyu} and \myref{eyd} that 
the $\ordnt{1}$ coefficients \myref{ooc}
satisfy
\beq
\arg D_u = \arg E_u = \arg \tilde C_u 
= \arg \tilde E_u = 0,
\eeq
\beq
\arg A_u = \arg C_u,
\eeq
with $\arg C_u, \arg B_d, \arg \tilde B_d$ and $\arg D_d$
independent parameters to be scanned over.
As in the previous case, we scan over the $\ordnt{1}$
magnitudes of the coefficients \myref{ooc}
using a Gaussian distribution, and plot
values of $V_{ub}(M_Z)$ for models which
satisfy all constraints in \myref{hidta} except
the one on $|V_{ub}|_{\Lambda_X}$.  The
results are given in Figure \ref{fig3}.
Comparing to Figure \ref{fig1}, it can
be seen that whether or not the mixings
$X^{1,2}$ are a source of phase makes little
difference.  Complex Xiggs vevs provide a
source of a KM phase and allow us to obtain
any value we like.

\newpage

\vspace{10pt}
\noindent {\it Case 3:  complex T-moduli}
\vspace{10pt}

\begin{figure}
\begin{center}
\includegraphics[height=5in,width=3.5in,angle=-90]{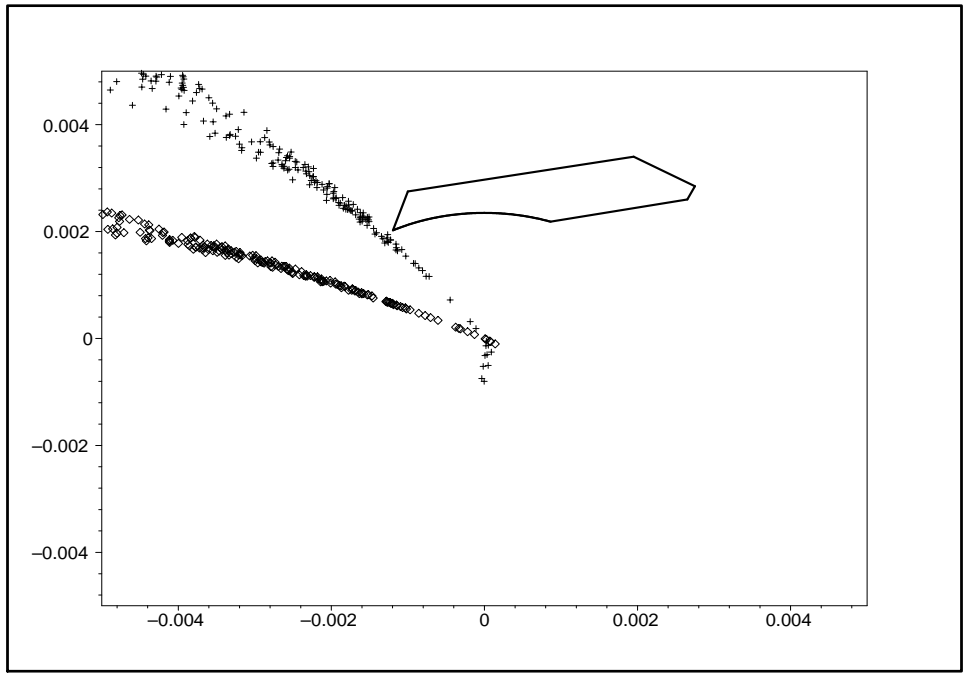}
\end{center}
\caption{$V_{ub}(M_Z)$ for positive parameters
except $\Gamma$.  The two nontrivial possibilities are
displayed:  $\Gamma= -\pi/3$ (crosses) and $\Gamma= -\pi/6$ (diamonds).}
\label{fig2}
\end{figure}

As discussed above, some of the 
T-moduli may stabilize at $e^{i\pi/6}$,
their other self-dual point under $SL(2,\Zbf)$.
If we keep the mixing matrices
$X^{1,2}$ and Xiggs vevs $\vev{Y_{1,2,3}^{\ell i}}$
complex, then the results are indistinguishable
from those of Fig.~\ref{fig1}.  To isolate
the effect of T-moduli sitting at the other
self-dual point, we have constrained the
coefficients $r_i,s_i$ in \myref{x12a}
and the Xiggs vevs $\vev{Y_{1,2,3}^{\ell i}}$
to be real and positive in what follows.
Next, referring to \myref{ytx}, \myref{eyu} and \myref{eyd}, 
we define
\beqa
\gamma_1 & \equiv & \arg h_u = \arg A_u = \arg C_u, \\
\gamma_2 & \equiv & \arg B_d = \arg
{\eta(T_c^1)^4 \chi_1(T_c^3) \over \eta(T_c^3)^4 \chi_0(T_c^3)}, \\
\gamma_3 & \equiv & \arg \tilde B_d = \arg D_d = \arg
{\eta(T_c^2)^4 \chi_1(T_c^3) \over \eta(T_c^3)^4 \chi_0(T_c^3)}, \\
\Gamma & \equiv & \gamma_1 - \gamma_2 + \gamma_3 .
\eeqa
It can be checked that the Yukawa matrices
\myref{eyu} and \myref{eyd} can be rephased such
that $\lambda^d$ has all positive entries and
\beq
\arg \lambda^u = \pmatrix{ \Gamma & 0 & \Gamma \cr
0 & 0 & 0 \cr 0 & 0 & 0 \cr}.
\eeq
This is easily implemented in a scan of
the parameters in \myref{ooc}
by requiring all of them to be positive
except $\arg A_u = \arg C_u = \Gamma$.
Using Table \ref{ecv} it is straightforward
to determine $\Gamma$.
We summarize the possible values in Table~\ref{table3}.
\begin{table}
\begin{center}
$$
\begin{array}{|l|l|l|l||l|l|l|l|} \hline
\arg T_c^1 & \arg T_c^2 & \arg T_c^3 & \Gamma &
\arg T_c^1 & \arg T_c^2 & \arg T_c^3 & \Gamma \\ \hline \hline
0     & 0     & 0     & 0      & 0     & \pi/6 & \pi/6 & -\pi/6 \\ \hline
\pi/6 & 0     & 0     & -\pi/6 & \pi/6 & 0     & \pi/6 & -\pi/6 \\ \hline
0     & \pi/6 & 0     & -\pi/6 & \pi/6 & \pi/6 & 0     & -\pi/3 \\ \hline
0     & 0     & \pi/6 & 0      & \pi/6 & \pi/6 & \pi/6 & -\pi/3 \\ \hline
\end{array}
$$
\end{center}
\caption{Phases from complex $T^I_c$.}
\label{table3}
\end{table}
The results of the scan are presented in 
Figure \ref{fig2}.  Once again,
these are values of $V_{ub}(M_Z)$ for models which
satisfy all constraints in \myref{hidta} except
the one on $|V_{ub}|_{\Lambda_X}$.

It can be seen that neither possibility 
is consistent with the
experimentally preferred region.  
This result only rules out T-moduli
as the sole source of CP violating
phases in the toy model considered here.
In another
model the powers of $\eta(T),\chi_0(T),\chi_1(T)$
would likely enter differently, since these
depend on the dimension of
a given nonrenormalizable coupling.  It
should also be noted that whereas
we have set the mixing matrices $X^{1,2}$ real,
they typically have a nontrivial
dependence on $\arg T^I_c$, since they are
determined at least in part by couplings
involving twisted fields.  Thus, the
results of Figure \ref{fig2} would likely
change if this part of the model were made
explicit.

%
%

\mysection{Conclusions}
\label{concl}
In this article we have discussed several
possible sources of CP violation in semi-realistic
heterotic orbifold models.
We have presented examples where CP violation
does not occur in spite of the presence of
phases, derived from complex string moduli vevs,
in renormalizable coupling coefficients.
However, it was described how nonrenormalizable
couplings give a significant contribution to
the effective quark Yukawa matrices when
an anomalous \ux\ is present and we
argued that this generically leads to
a nontrivial KM phase.  

In order to make a detailed analysis of
models with nonrenormalizable couplings,
we introduced modular covariant nonrenormalizable
superpotential couplings.  It was explained
why it is difficult to obtain the
{\it bona fide} effective coupling coefficients from
conformal field theory techniques.
It was also pointed out that higher
order terms in the K\"ahler potential
should be important in cases
where an anomalous \ux\ is present.
These theoretical uncertainties represent
a significant stumbling block to
further progress in string-derived
effective supergravity models and
it is hoped that they will be resolved
at some point in the near future.

The KM phase was determined explicitly in a toy
model inspired by three-generation
heterotic $Z_3$ orbifold constructions.
Though target space modular invariance and
orbifold selection rules greatly restrict the
coupling coefficients of nonrenormalizable
couplings, we found it possible to obtain
viable Yukawa couplings for quarks
by adjusting the vevs of Xiggses.  This
result highlights the necessity of understanding
how D-moduli flat directions are lifted
in a given model.  In principle the
Xiggs vevs should be determined by the
mechanisms which lift these flat directions.
This would eliminate our ability to tune
the scalar background to our liking and
would in most cases probably render
the quark phenomenology inconsistent with
low energy data.  We are currently
investigating this issue and hope to
report on it in a future publication.

\vspace{20pt}
\noindent {\bf \large Acknowledgements}
\vspace{10pt}

The author would like to thank Prof.~Mary K.~Gaillard
for innumerable discussions and helpful comments
during the development of the work contained
here.  I am particularly indebted to her
for showing me how the Green-Schwarz
cancellation of the \ux\ anomaly works in the linear multiplet
formulation and have borrowed heavily from
written communications with her in preparing the
Appendix.  I would also like to thank
Brent Nelson for useful comments.
This work was supported in part by the
Director, Office of Science, Office of High Energy and Nuclear
Physics, Division of High Energy Physics of the U.S. Department of
Energy under Contract DE-AC03-76SF00098 and in part by the National
Science Foundation under grant PHY-95-14797.

\myappendix

\mysection{Charge normalization}
\label{normsub}
In the FIQS model, unconventional normalizations for
the $U(1)$ charges have been chosen to keep the tables
of charges simple and amenable to computer
assisted analysis.  The generator $Q_X^I$ acting on the
$E_8 \times E_8$ root torus is given by
\beq
Q_X = 6 (0, 0, 0, 0, 0, 0, 0, 0 ; 1, -1, 1, 0, 0, 0, 0, 0).
\eeq
The affine level of a $U(1)$ group may be defined \cite{kobay:97a}
as
\beq
k_Q = 2 \sum_{I=1}^{16} (Q^I)^2.
\eeq
With this convention, the FIQS normalization
gives $k_{Q_X} = 6^3$.  To go to a normalization where the coupling
constant for the \ux\ group will be the universal coupling at the
string scale, we must rescale the generator $Q_X \to Q_X'$ so
that $k_{Q_X'} = 1$.  Then
\beq
Q_X' = {1 \over 6 \sqrt{6}} Q_X .
\eeq
Since the original normalization satisfied\footnote{The
nonabelian generators $T^a$ are normalized such
that $\tr \, T^a T^a = 1/2$ for a fundamental
representation of $SU(N)$.}
\beq
\tr Q_X^3 = 27 \cdot \tr Q_X
= 27 \cdot 24 \cdot \tr T^a T^a Q_X = 27 \cdot 24 \cdot 54 ,
\eeq
it can be checked that the rescaled generator satisfies
\beq
24 \tr (T^a T^a Q_X') = 8 \tr {Q_X'}^3 = \tr Q_X' = 36 \sqrt{6},
\label{canonqx}
\eeq
as required by anomaly matching \cite{kobay:97a}.
Indeed, if the \ux\ vector superfield $V_X$ is shifted
by $\delta V_X = (1/2)(\Lambda + \bar \Lambda)$,
then the resulting anomalous transformation of
the lagrangian is
\beq
\delta {\cal L} = {1 \over 16 \pi^2} \sum_a \tr (T^a T^a Q_X')
\left[ \Real \lambda F^a \cdot F^a + \Imag \lambda F^a \cdot \tilde F^a
\right] + \cdots
\eeq
where $\lambda = \Lambda |$.  We introduce
our counterterm\footnote{We work in K\"ahler
superspace \cite{bggm} and use the linear
multiplet formulation where $V$ is a real superfield
which satisfies {\it modified linearity conditions}
and contains the dilaton $\ell$ as its lowest
component \cite{abgg,binet:97a}.} as
\beq
{\cal L}_{GS,V_X} = \delta_X \int E V V_X
\label{gsvx}
\eeq
from which it follows that under the shift in $V_X$
\beq
\delta {\cal L}_{GS,V_X} = {\delta_X \over 2} \int E V (\Lambda + \bar \Lambda)
\eeq
which when we go to components yields
\beq
\delta {\cal L}_{GS,V_X} = - {\delta_X \over 8}
\sum_a \left( \Real \lambda F^a \cdot F^a
+ \Imag \lambda F^a \cdot \tilde F^a \right) + \cdots
\eeq
The anomaly is cancelled if we choose
\beq
\delta_X = {1 \over 2 \pi^2} \tr T^a T^a Q_X'.
\eeq
When combined with other terms in the lagrangian,
the component form of \myref{gsvx} gives
\beq
D_X = \sum_i q_X^i \hat K_i \phi^i + {\delta_X \over 2} \ell
\equiv \sum_i q_X^i \hat K_i \phi^i + \xi.
\eeq
From this, we see that the FI term $\xi$ is given
by
\beq
\xi = (2 \ell) {\delta_X \over 4} = {2 \ell \over 8 \pi^2}
\tr T^a T^a Q_X'.
\eeq
With the \ux\ generator chosen such that $k_{Q_X'}=1$,
equation \myref{canonqx} gives
\beq
\xi = {2 \ell \over 192 \pi^2} \tr Q_X' ,
\eeq
which may be recognized as the form typically quoted
in the literature once it is realized that if we neglect
nonperturbative corrections to the K\"ahler potential
of the dilaton $\ell$, the universal coupling constant
at the string scale is given by $g^2 = 2 \ell$.
In the FIQS normalization, the FI term is given by
\beq
\xi = {2 \ell \over 192 \pi^2} {1 \over 6 \sqrt{6}} \tr Q_X
\eeq
which gives a significantly smaller number than if
we had not accounted for the unconventional
normalization of the \ux\ charge.  In the FIQS model
$\tr Q_X = 1296$, yielding
\beq
\xi \approx 2\ell \times 4.7 \times 10^{-2} \sim 5 \times 10^{-2},
\eeq
where we have used $2\ell \approx g^2$,
and $0.5 \lappeq g^2 \lappeq 1$.
The scale of \ux\ breaking is given by $\Lambda_X
\sim \sqrt{\xi} \sim 0.22 \, m_P \approx 5 \times 10^{17}$ GeV
$\approx \Lambda_s$, the string scale.

One must also take proper account of charge normalization for
the SM hypercharge, as was pointed out in ref.~\cite{font:90a}.
For example, in the FIQS model $k_Y = 11/3$.
Then the charge generator which will have the unified coupling
at the string scale is $Y' = \sqrt{3/11} \; Y$.
This is to be compared with the
$G_{GUT} \supseteq SU(5)$ relative factor of
$\sqrt{3/5}$.  Thus, the boundary value of the properly
normalized hypercharge coupling $g'$
at the electroweak scale in the FIQS model is
related to the one usually used in GUT-inspired
renormalization group evolution of the couplings
in the MSSM by
\beq
\left. g'( \mtxt{FIQS} ) \right|_{M_Z}
= \sqrt{11/5} \; \left. g'( \mtxt{MSSM} ) \right|_{M_Z}.
\eeq
This clearly does violence to unification of the
couplings.  In short, it is necessary to include
hypercharge normalization among the criteria to
be checked when searching for viable string-derived
models.

\baselineskip=16pt

%
%

\end{document}